\documentclass[12pt,preprint]{aastex}

\newcommand{\sollum} {\hbox{$L_{\odot}$}}
\newcommand{\solmass} {\hbox{$M_{\odot}$}}

\shorttitle{Maser Monitoring}
\shortauthors{Braatz et al.}

\begin{document}

\title{A Survey for H$_2$O Megamasers III. Monitoring Water Vapor Masers in 
Active Galaxies}

\author{J. A. Braatz}
\affil{NRAO\altaffilmark{1}, P.O. Box 2, Green Bank, WV 24944}
\altaffiltext{1}{The National Radio Astronomy Observatory is a facility of
the National Science Foundation operated under cooperative agreement with
Associated Universities, Inc.}
\author{A. S. Wilson\altaffilmark{2}}
\affil{Department of Astronomy, University of Maryland, College Park, MD 20742}
\altaffiltext{2}{Adjunct Astronomer, Space Telescope Science Institute,
Baltimore, MD 21218}
\author{C. Henkel}
\affil{Max-Planck-Institut f\"{u}r Radioastronomie, Auf dem H\"{u}gel 69,
D-53121 Bonn, Germany}
\author{R. Gough and M. Sinclair}
\affil{CSIRO Australia Telescope National Facility, P.O. Box 76, Epping,
NSW 2121, Australia }

\begin{abstract}
We present single-dish monitoring of the spectra of 13 extragalactic water
megamasers taken over a period of 9 years and a single epoch of sensitive
spectra for 7 others.
The primary motivation is a search for drifting line velocities
analogous to those of the systemic features in NGC 4258, which are known to
result from centripetal acceleration of gas in an edge-on, sub-parsec molecular
disk.  We detect a velocity drift analogous to that in NGC 4258 in only
one source, NGC 2639.  Another, the maser
source in NGC 1052, exhibits erratic changes in its broad maser
profile over time.  Narrow maser features in all of the other
disk galaxies discussed here either remain essentially constant in velocity
over the monitoring period or are sufficiently weak or variable in intensity
that individual
features cannot be traced reliably from one epoch to the next.
In the context of a circumnuclear, molecular disk model,
our results suggest that either (a) the maser lines seen are systemic features
subject to a much smaller acceleration than present in NGC 4258, presumably
because the gas is farther from the nuclear black hole,
or (b) we are detecting ``satellite'' lines for which the
acceleration is in the plane of the sky.

Our data include the first K-band science observations taken with the new 100 m
Green Bank Telescope (GBT).  The GBT data were taken during testing and
commissioning of several new components and so are subject to some limitations;
nevertheless, they are in most cases the most sensitive H$_2$O spectra ever
taken for each source and cover 800 MHz ($\simeq$ 10800 km s$^{-1}$) of 
bandwidth.  Many new maser features are detected in these observations.
Our data also include a tentative and a clear detection of the megamaser in 
NGC 6240 at epochs a year and a few months, respectively, prior to the
detections reported by \citet{hagiwara02a} and \citet{nakai02}.

We also report a search for water vapor masers towards the nuclei of 58 highly
inclined (i $>$ 80$^\circ$), nearby galaxies.  These sources were selected
to investigate the tendency that H$_2$O megamasers favor inclined galaxies.
None were detected, confirming that megamasers are associated exclusively with 
active galactic nuclei.
\end{abstract}

\keywords{galaxies: active --- galaxies: nuclei --- galaxies: Seyfert ---
ISM: molecules --- masers --- radio lines: galaxies}

\section{Introduction}

VLBI mapping of water masers towards the nucleus of the low luminosity active 
galaxy NGC 4258 reveals a thin, edge-on disk 0.28 pc in radius
\citep{herrnstein99}.
The maser spectrum has a characteristic profile with three clusters of
features: the strongest cluster is within 50 km s$^{-1}$ of the 
systemic velocity of the galaxy (``systemic features'') while the others
are red- and blue-shifted by $\simeq$ 1000 km s$^{-1}$ w.r.t systemic 
(``high velocity features'').  The three clusters are associated with three 
loci in the rotating gas disk; each locus corresponds to a peak in the 
line-of-sight velocity coherence in the disk, thus 
producing the longest path lengths for maser amplification.  The systemic 
features are associated with gas on the near side of the disk, close to the
line of sight to the nucleus and radio continuum jets.  The
high velocity features are found along the ``edges'' of the disk, where the 
rotational velocity vector points approximately along our line of sight.

VLBA observations have been performed on those other megamaser galaxies which 
are bright enough for self-calibration on the maser features themselves 
(e.g. NGC 4945 -- \citep{greenhill97c}, NGC 1068 -- \citep{greenhillgwinn97}, 
NGC 3079 -- \citep{trotter98}, IC 2560 -- \citep{ishihara01}).  The results 
suggest that a gas disk is usually present but, in those sources which contain 
one, the dynamical structure of the masing disk is not as
clean as in NGC 4258 (i.e. the disk is not in strictly 
Keplerian rotation, or maser emission is detected away from the disk).

The maser spectrum of NGC 4258 changes with time in a fashion which can be
understood in the context of the rotating nuclear disk model.  Narrow 
lines among the systemic features drift redward through a $\simeq$ 100
km s$^{-1}$ wide ``envelope'' at a rate of $\simeq$ 9 km s$^{-1}$ yr$^{-1}$
\citep{haschick94, greenhill95}.  This velocity drift reflects the 
gravitational acceleration of gas on the near side of the disk
as it orbits the central black hole.  The acceleration vector of such
gas points directly away from the observer.
For the high velocity features, the acceleration is directed in the plane of 
the sky and the maser lines remain at almost constant velocities, although 
complexities in the gaseous disk and deviations in the positions of the masers 
from the midline may lead to small velocity changes
($<$ 0.8 km s$^{-1}$ yr$^{-1}$) in these components \citep{bragg00}.

In this paper we investigate the time dependent behavior of water vapor 
masers in 13 galaxies and present recent sensitive spectra of 7 additional 
galaxies.  We also report a search for water vapor masers towards the nuclei
of 58 highly inclined, nearby galaxies.

\section{Observations}

We have monitored the 22 GHz water maser line of 13 galaxies between October 
1993 and November 2002.  Observations were made at Effelsberg using the 100 m
telescope, at Parkes using the 64 m telescope, and at Green Bank using the 
140-foot telescope and the 100 m Green Bank Telescope (GBT).
Table 1 lists the dates and locations of 
the observing runs.  The velocity scale recorded by the on-line software in 
Effelsberg uses the optical convention while the spectral data reduction
program CLASS expects the radio convention, so a correction must be made to 
those data.  Afterwards, the velocities of individual maser features were 
checked between the four telescopes and found to be consistent 
to better than 1 km s$^{-1}$.  Observing conditions varied drastically, and 
uncertainties in absolute flux densities are 10\% to 30\%.  Thus, while our data
are excellent for investigating time dependent velocity variations, uncertain 
flux calibration and sometimes infrequent sampling make
detailed flux variability studies difficult.

The GBT data shown here were taken during commissioning and testing of the 
telescope, and are part of the first K-band experiment performed with the GBT.
An 800 MHz bandwidth mode of the spectrometer and the active surface 
were being tested for the first time during this experiment as well.  The 
active surface was running in an open loop mode and was programmed to counter
the effects of gravitational deformations as determined by a finite element 
model.  Aperture efficiencies ranged between 50\% and 60\% over elevation 
ranges from 15$^\circ$ to 85$^\circ$ with the active surface on, roughly a 
factor of two improvement over the efficiency with the active surface off.
Because of the test nature of these observations, not all components of the 
GBT were available.  We used only one polarization and only half of the 
available integrations due to testing of a new mode of the spectrometer.  
The effective integration times are therefore shorter with the GBT 
(10 minutes -- 1 hour) than the other telescopes; nevertheless, the GBT 
spectra represent, in most cases, the most sensitive taken on these galaxies.
System temperatures for the GBT data ranged from 37K to 60K.

For the Effelsberg and Green Bank 140 ft observations, a pointing calibrator 
was observed prior to each source, resulting in a pointing accuracy of 
about 10$\arcsec$.
Pointing at Parkes, using the Master Equatorial System, has similar accuracy.
Local pointing calibrators were used with the GBT observations as well.  The
pointing uncertainty there was 6$\arcsec$ -- 8$\arcsec$ before applying the 
local
corrections and $\lesssim$ 2$\arcsec$ afterwards.  All observations prior to
October 1999 utilized position switching to subtract the background.
The autocorrelators were typically used in a wide bandwidth mode (50 MHz at 
Effelsberg, 64 MHz at Parkes, and 40 MHz for the Green Bank 140 ft) to
maximize the velocity coverage.  Observations since October 1999 at Effelsberg 
made use of the new dual beam system and new autocorrelator, running in an 
8 $\times$ 40 MHz mode with polarizations overlapped, resulting in $\sim$ 150 
MHz effective coverage.  The GBT data were also taken with a dual beam 
system using beam switching to remove the background.
Baseline curves were removed from the spectra by fitting a polynomial function,
and a sinusoidal term as well for a few 140 ft observations.  A second or 
third order polynomial across the 40 MHz -- 64 MHz bandwidths generally 
provided a satisfactory description of the baseline.
With the 140 ft, the focus was modulated and a double switching mode was 
used, with excellent baselines resulting.  When fitting a baseline to the
full 800 MHz of the GBT spectra, a higher order polynomial was generally 
required.  For the purposes of this program, baseline waviness introduces a 
problem only when the width of the emission line feature is a
significant fraction of the wavelength of the baseline artifacts.  NGC 1052 
and TXS 2226-184 are thus most seriously affected.  The K-band receiver used 
with the Green Bank 140 ft telescope was upgraded after November 1994, providing
a significant improvement in sensitivity.  Additional observing details are 
given in \citet{braatz96}.

The velocities in this paper follow the optical definition and are calculated
with respect to the Local Standard of Rest (LSR).  Galaxy recession velocities 
quoted here are taken from the HI velocities in \citet{deV91}
and are converted to LSR for comparison with the maser spectra.  Data were
reduced using the software packages UNIPOPS (NRAO 140 ft), CLASS 
(Effelsberg 100 m), SPC (Parkes) and AIPS++ (GBT).

\section{Line Widths and Intensities}

The width of observed megamaser lines ranges from $\lesssim$ 1 km s$^{-1}$
(e.g. NGC 5506 and the Circinus Galaxy) to several hundred km s$^{-1}$ 
(NGC 1052 and TXS 2226-184).  For some sources (e.g. NGC 4945) a broad maser 
feature is resolved into a group of closely spaced narrow spikes, but not so 
for NGC 1052 and TXS 2226-184.  The width of the maser emission feature in 
NGC 1052 varies (Figure 1), and although the baselines were occasionally 
difficult to determine, variations in the maser profile are clearly
seen.

One of the characteristic features of cosmic masers, both Galactic and
extragalactic, is rapid variability of their observed intensities. 
The timescale and amount of variability, though, differs drastically for 
the individual sources we have 
monitored.  The maser in NGC 1386, for example, has remained mostly unchanged
during our 8 years of monitoring, while the maser in the Circinus Galaxy
has changed its profile entirely between each observation.  Indeed, variability
in the Circinus maser is reported on timescales as short as several minutes
\citep{greenhill97a}, possibly due to interstellar scintillation.  Variability
revealed by the present observations, though, is likely to have an 
intrinsic component as well (e.g.  \citet{kartje99}).
We have detected strong and short-lived (several month timescale) flares.
In some cases, flares seem to arise from previously dimmer features, as in 
NGC 5506 starting March 1995 and IC 2560, which was seen to be bright in
August 1997.  In other cases, flares rise up from below the noise level.
The sensitivity improvements due to the upgrades of the Effelsberg telescope
and the construction of the GBT reveal a wealth of fainter features, as
highlighted by the GBT spectrum of NGC 5506 (Figure 2).

Intrinsic variability in maser intensity might result from either a change 
in the pumping source or a change in the gain path.  Some AGN are known to 
have rapidly variable X-ray sources which might ultimately provide the pumping
energy.  X-ray variability is stronger for Seyfert 1 galaxies, in which 
the X-ray source is thought to be seen directly, than for Seyfert 2s, in which 
a scattered or intrinsically extended component of the X-rays dominates at
soft X-ray energies.  The difference is believed to be purely one
of viewing angle (e.g. \citet{antonucci93}) so even though water megamasers
are detected in only Seyfert 2s and LINERs,
the relevant X-ray variability timescale is that of Seyfert 1s.  
NGC 1068 is an interesting case when considering this type of behavior.  Over 
the past decade, several prominent maser features in NGC 1068 have disappeared 
and ``new'' ones appeared at later times \citep{baan96} having the 
same velocity.  The reappearance suggests that each of
these maser features is associated with a molecular gas complex which does
not disappear over the monitoring interval, so the variability might be
partly due to a changing pumping source.  Flares from previously dimmer
features, seen in several instances in Figure 1, may have a similar origin.

In NGC 4258, at least some of the maser variability is due to the orbital 
motion of molecular clouds.  The maser associated with a given systemic cloud 
is only detectable while that cloud is in a small range of its orbit on the
near side of the disk.
However, the timescale for clouds to pass through this range ($\sim$ 10 
years) is much longer than the observed variability of individual features.
More likely, the strength and structure of maser features in this source and 
most others depend strongly on the geometry and orientation of orbiting 
molecular complexes.  

\section{GBT Results}

Because faint features evident in the GBT data are poorly represented on the 
flux scales in Figure 1, we reproduce a GBT spectrum (Figure 2) of 10 of the
sources monitored as well as 8 others visible in the 
northern sky.  Although the GBT spectrometer covered 800 MHz of simultaneous
bandwidth, no maser lines were detected beyond $\sim$ 80 MHz from the systemic
velocity in any of the sources.  The spectra in Figure 2 show 1600 km s$^{-1}$
coverage for all sources except NGC 4258, for which the coverage is expanded
to 2600 km s$^{-1}$ to show the high velocity features, and Mrk 1419 (see
\citet{henkel02}), which covers 2000 km s$^{-1}$.  The insets to the
spectra for NGC 1386 and NGC 4258 are correctly matched to the velocity scale
but show an expanded (and offset) flux scale to emphasize the detections of 
faint features.

\section{Comments on Individual Sources}

The maser in ESO 103-G35 was detected
only once \citep{braatz96} in June 1994, and will not be discussed here.
Mrk 348 \citep{falcke00a} was not detected in H$_2$O during the GBT test
observations and has not been monitored during this program.
Masers in NGC 3735 \citep{greenhill97b} and NGC 5793 \citep{hagiwara97}
have also not been monitored by us but were observed with the GBT, and
the resulting spectra are shown in Figure 2.  No significant new
features are detected in these sources, but the signal-to-noise is 
better than in previous data.  Comments on other maser-detected galaxies follow.

Flux densities in Figure 1 are calibrated in Jy but the spectra are shifted 
from each other for clarity.  Rather than showing the full spectra, a 
limited velocity 
window is shown for each source to facilitate comparisons of
the maser velocities from one epoch to the next.  No maser components were
detected outside of the windows shown, except where noted in Mrk 1210.
Line velocities used to measure maser drift rates have been determined by 
fitting Gaussian curves to each line, where possible.

\subsection{Mrk 1}
A single faint ($\simeq$ 0.1 Jy) maser feature near 4868 km s$^{-1}$
has been seen over 8 years of monitoring, and shows no detectable 
velocity drift during that period (--0.15 $\pm$ 0.13 km s$^{-1}$ yr$^{-1}$).
This maser line is redshifted from the 21 cm HI velocity 
(4825 km s$^{-1}$) by 43 km s$^{-1}$.  The more sensitive spectra 
of this source reveal that
the line is the strongest feature of a cluster of maser features extending
some 60 km s$^{-1}$ and centered near 4860 km s$^{-1}$.  Weaker blueshifted 
features are evident near 4583 and 4620 km s$^{-1}$ and likewise show no 
detectable drift during the monitoring period 
(0.00 $\pm$ 0.16 km s$^{-1}$ yr$^{-1}$ and 
0.02 $\pm$ 0.17 km s$^{-1}$ yr$^{-1}$, respectively).

\subsection{NGC 1052}
NGC 1052 is the only elliptical galaxy detected in H$_2$O,
and the maser in this source is clearly unlike those in disk galaxies
(but, see also the discussion for TXS 2226-184).
The broad maser in NGC 1052 seemed to drift redward between March 1994 and 
December 1995, and subsequently showed complex structure, with an apparent
bifurcation of the broad peak evident in the November 1997 spectrum.  Several
of the spectra, especially those in November 1994 and February 1998, are
adversely affected by baseline structure.  Nevertheless, the maser
profile clearly changed during the monitoring period.

VLBI spectral line observations in November 1995 \citep{claussen98}
show that the maser in NGC 1052 arises primarily from two clumps of gas
having distinct but overlapping velocities.  The two clumps align parallel to 
the long known nuclear radio jet.  The maser gas may be entrained along the 
jet or perhaps a disk lies in front of one side of the jet, in which case the 
masers may originate in gas amplifying the background continuum emission of the
jet (in NGC 4258 the maser disk projects perpendicular to the radio
continuum jet).  It 
is not clear whether the broad line emission represents
gas being accelerated by the jet or whether it is unrelated to
the gas dynamics.  It is possible that the profile changes result simply from
gas at different line-of-sight velocities being lit up with maser emission
at different times.

The GBT spectrum shows that in April 2002, after a lengthy gap from the
previous spectrum, the maser had dimmed and possibly broadened.
That the emission lacks narrow lines is counter to the typical maser profile,
and the reason for the smooth, broad profile remains unknown.

\subsection{NGC 1068}

The maser source in NGC 1068 has been studied extensively by \citet{baan96}
and \citet{gallimore01}, and has not been part of our long-term monitoring
program.  We do present a GBT spectrum in Figure 2.

\subsection{NGC 1386}
This nearby Seyfert 2 galaxy has two prominent narrow features near 967 and
977 km s$^{-1}$ and a broad feature at $\simeq$ 860 km s$^{-1}$, near the 
galaxy's systemic velocity of 847 km s$^{-1}$.  The drift in the narrow
features is measured to be --0.07 $\pm$ 0.14 km s$^{-1}$ yr$^{-1}$ (967) and
--0.12 $\pm$ 0.14 km s$^{-1}$ yr$^{-1}$ (977).  Individual features in the
broad systemic component have not been traceable from one epoch to the
next because of the signal-to-noise limitations and blending of the lines,
but it is clear that future studies making use of the GBT will be able to track
such features.  The GBT spectrum also reveals several blueshifted components
near 600 km s$^{-1}$.  It is possible that the features near 600 and 970 
km s$^{-1}$ are analogous to the high velocity features, and those near
860 km s$^{-1}$ to the systemic features, in NGC 4258.
Except for flux variability in the narrow redshifted 
features, the maser profile in this galaxy appears remarkably unchanged
over the monitoring period.

\subsection{Mrk 1210}

Narrow maser features were first detected near velocities 4214 and 4247 
km s$^{-1}$, redshifted from the systemic velocity of this galaxy 
(4032 km s$^{-1}$).  The drifts in these lines are measured to be
--0.18 $\pm$ 0.11 km s$^{-1}$ yr$^{-1}$ and
--0.12 $\pm$ 0.16 km s$^{-1}$ yr$^{-1}$, respectively.
If the blueshifted line at $\sim$ 4091 km s$^{-1}$ detected in April 2002 is 
associated with the line at $\sim$ 4094 during 1998, then it had a blueward 
velocity drift of --0.93 $\pm$ 0.30 km s$^{-1}$ yr$^{-1}$.
Another line is found at 4226 km s$^{-1}$ and has no detectable drift
(--0.17 $\pm$ 0.42 km s$^{-1}$ yr$^{-1}$).
The GBT spectrum (Figure 2) shows a blueshifted feature centered near 3780
km s$^{-1}$.  This feature was also detected (but not shown in Figure 1 in 
order to see other features clearly) in Effelsberg spectra from March 2000 and
January 2001.  The measured velocity drift for this feature is
0.81 $\pm$ 0.57 km s$^{-1}$ yr$^{-1}$.

\subsection{NGC 2639}

\citet{wilson95} discussed the initial results of the monitoring
program on NGC 2639, arguing that the galaxy and its maser share a number
of similarities with NGC 4258.  Most significantly, the strongest maser feature
detected between October 1993 and March 1995 drifted redward at a rate of 6.6
km s$^{-1}$ yr$^{-1}$.  The galaxy was observed several times with the 
Effelsberg 100 m 
telescope during the next two years but the maser was not detected, apparently
due to a combination of poor weather conditions during the observing and 
diminishing maser brightness.  The maser was detected again in 1997 and 1998.
The drift of the line initially seen at 3307 km s$^{-1}$ over the
period Apr 1997 -- Feb 1998 is
6.5 $\pm$ 1.6 km s$^{-1}$ yr$^{-1}$, consistent with the results of
\citet{wilson95}.
A pair of 14 mJy lines was detected with the GBT in April 2002 (Figure 2).

The maser emission was observed to be much brighter in April 1997.  At that 
time, broad emission was seen between about 3220 and 3310 km s$^{-1}$, which is 
slightly to the blue of the systemic HI velocity, 3335 km s$^{-1}$.  
If the masers detected in April 1997 delimit the envelope of systemic maser 
features (see also the February 1994 spectrum), analogous to the masers in 
NGC 4258, then the disappearance of the maser feature tracked between 
October 1993 and March 1995 may be understood in terms of the masing gas 
drifting beyond the envelope of observable systemic features (cf. 
\citet{greenhill95} for NGC 4258).

\subsection{Mrk 1419}

\citet{henkel02} report the discovery of a triply-peaked H$_2$O maser
towards the nucleus of Mrk 1419.  The authors find that the maser source 
has properties similar to those seen in NGC 4258, including an acceleration
in the ``systemic'' masers (2.8 $\pm$ 0.5 km s$^{-1}$ yr$^{-1}$) and
no acceleration in the ``high velocity'' lines.  In Figure 2 we present a 
sensitive spectrum of Mrk 1419 taken with the GBT in April 2002.  No features
were detected outside of the velocity range reported by \citet{henkel02}.

\subsection{NGC 3079}

The water maser source in NGC 3079 has been studied extensively by 
\citet{baan96} and \citet{hagiwara02b}
and has not been part of our monitoring program.  However, we present a GBT
spectrum of this maser (Figure 2) which is shown with a flux range chosen to
emphasize the weak features.  Our spectrum confirms the redshifted structure
detected by \citet{hagiwara02b} as well as the narrow gap at the systemic 
velocity, though the flux in the April 2002 spectrum does not drop to zero at 
the systemic velocity, as in their December 2001 data.

\subsection{IC 2560}

Observations of the maser in this galaxy clearly benefit from the high 
sensitivity
of the GBT.  Since the discovery of the maser in June 1994 \citep{braatz96} 
with the Parkes telescope, this source has been observed several more times at 
Parkes and with the NRAO 140 ft.
As with NGC 1386, monitoring of this source is sparse because
it is too far south for observation with the Effelsberg telescope.  A flare
in August 1997 resulted in roughly a factor of two increase over the typical
flux, but the flare had subsided by December 1997.
Although our data show no velocity drift over the 4 years of monitoring, it is
clear that from 1994 to 1998 we were not distinguishing individual components in
the systemic clump.  \citet{ishihara01} have observed this source with both
the NRO 45 m and Parkes 64 m telescopes, and have attained a greater 
sensitivity than the 1994 -- 1998 spectra shown here.  \citet{ishihara01} 
resolve 4 -- 6 systemic features at each epoch and find a redward velocity 
drift in each
of 2.6 km s$^{-1}$ yr$^{-1}$.  They also report the discovery of numerous
high velocity components, which are confirmed by our April 2002 spectrum
(Figure 2).  Note that the velocity scale appears different between the
spectra here and those presented by \citet{ishihara01} because of the
difference between the optical and radio velocity conventions.

\subsection{NGC 4258}

The maser in NGC 4258 has been monitored by \citet{haschick94}, 
\citet{greenhill95} and \citet{bragg00}.  We have not monitored this source as
part of our program, but we do present a GBT spectrum in Figure 2.  In the 
triply-peaked structure of the maser in NGC 4258, the systemic masers are
clearly the brightest.  A similar relative strength of components is
apparent in IC 2560.  However, if the negligible velocity drifts in the masers
of other detected galaxies indicate that they are analogous to the 
high velocity lines in NGC 4258, then the high velocity lines are
dominant in these galaxies.

\subsection{M51}

The kilomaser in the mildly active nucleus of M51 was first detected by 
\citet{ho87} and has been studied more recently by \citet{hagiwara01}.
This maser has not been a part of our monitoring program, but we present a 
sensitive GBT spectrum in Figure 2.  The blueshifted feature evident in
the January 2001 spectrum \citep{hagiwara01} had diminished by the April 2002
spectrum shown here.  A new feature appears at 546 km s$^{-1}$ and two equal
intensity ($\sim$ 0.1 Jy) components are found at 564 and 567 km s$^{-1}$.

\subsection{NGC 4945}

In NGC 4945 we detected maser emission near the systemic velocity of 556 
km s$^{-1}$ and redshifted from systemic, between about 630 and 740 
km s$^{-1}$.  \citet{greenhill97c} mapped the features in  this source using 
the subset of VLBA antennas to which NGC 4945 is accessible.
They detect blueshifted emission not seen in our spectra (and not shown in the
velocity range displayed in Figure 1).  \citet{greenhill97c} find an elongated
maser distribution possibly revealing a disk.  The systemic and redshifted 
features are separated by $\sim$ 0.15 mas (0.3 pc).  One might then expect 
that the maser features near 556 km s$^{-1}$ are analogous to the systemic 
features of NGC 4258.  Similarly, the cluster of features to the red 
might be associated with the ``edge'' of the disk (i.e. be ``high velocity
features'').  Although Gaussian fits to
individual components are difficult in the redshifted features, it is possible 
to trace local maser peaks throughout the monitoring period.  Drifts are
negligible, as expected for high velocity lines.
We measure drifts of 0.20 $\pm$ 0.31 , 0.16 $\pm$ 0.31, and 0.25 $\pm$ 
0.31 km s$^{-1}$ yr$^{-1}$ for peaks at 705, 716 and 732 km s$^{-1}$.  The 
equality of the error estimates results from using channel widths as error
estimates for each individual epoch.  The strongest systemic line 
also shows no discernible drift.
By resolving the strong systemic peak into two Gaussian
components at each epoch, we find that those lines (centered at 553 and 555
km s$^{-1}$) drift by 0.07 $\pm$ 0.35 and --0.07 $\pm$ 0.46
km s$^{-1}$ yr$^{-1}$.  
Assuming the disk model fits, a constraint on the black hole mass can be made.
Taking a nominal upper limit for the acceleration of $<$ 0.1 km s$^{-1}$ 
yr$^{-1}$ and rotation of 130 km s$^{-1}$, the central black hole mass 
must be $>$ 7 $\times$ 10$^5$ $\solmass$.

\subsection{NGC 5347}

The maser in this galaxy was tentatively detected in October 1993 and
confirmed by several successive observations, which show a group of
weak masers spanning $\sim$ 100 km s$^{-1}$ centered near the recession 
velocity of the galaxy, 2398 km s$^{-1}$.  Although there are hints of redward
drifts in the recent spectra, a more finely sampled set of observations is
required for confirmation.  The GBT spectrum reveals multiple distinguishable 
components in the systemic clump but no unambiguous detections of high 
velocity features.

\subsection{The Circinus Galaxy}

Because of the complexity of the maser emission in the Circinus Galaxy it has 
been divided into two segments for Figure 1, each covering a different velocity 
range.  Maser emission is distributed over a range of 400 km s$^{-1}$ 
bracketing the systemic velocity, 433 km s$^{-1}$.  \citet{greenhill97a} have
shown that the flux of the maser lines in this galaxy vary on timescales of
minutes to hours.  They suggest that such intraday flux variations may 
be caused by interstellar scintillation from the ISM of our Galaxy.  As 
evident in Figure 1, strong variability exists over longer timescales as well 
(note that 
each of the Circinus spectra shown is the result of a $\sim$ 2 hour 
integration, thereby smoothing over much of the 5 -- 10 minute timescale 
variability).  While several narrow features do appear to persist from one 
epoch to the next, most change so much that it is difficult to track velocity
drifts.  Unlike the narrow lines, though, the broad feature between 325 and 400 
km s$^{-1}$ persists throughout the series of spectra.  Uncertainty in
the baseline in October
1995 may account for the weakened broad feature for that epoch.  Several
clumps of maser features seem to persist as well, particularly in the
redshifted segment where features grouped near 512, 560, and 600 km s$^{-1}$
exist throughout the 4.3 year period, though the detailed structure
within those clumps changes.

\subsection{NGC 5506}

The maser in the edge-on Seyfert 1.9 galaxy NGC 5506 has been marked by 
several flares which have lasted roughly a year, and then disappeared.
A flare near 1730 km s$^{-1}$ was detected during 1993 -- 1994 and a
feature near 1800 km s$^{-1}$ flared spectacularly through
1995.  The numbers in Figure 1 printed adjacent to the flaring features 
indicate the peak intensity in Jy for the maser spike in cases where the
plots overlap.  Velocity drifts of lines at 1731 and 1799 km s$^{-1}$ are
measured to be 0.36 $\pm$ 0.71 km s$^{-1}$ yr$^{-1}$ and 0.00 $\pm$
0.29 km s$^{-1}$ yr$^{-1}$, respectively.  The isolation and narrow profiles
of these lines inspire confidence that each originates from a single,
persistent feature rather than a blend.
The GBT spectrum (Figure 2) shows a group of weak, narrow features centered
near the systemic velocity of the galaxy, 1824 km s$^{-1}$.

\subsection{NGC 6240}

Although observed many times and tentatively detected by \citet{henkel84}, 
H$_2$O maser emission was not conclusively detected in NGC 6240 until 2001.  
Both \citet{hagiwara02a} and \citet{nakai02} reported the detection in late
spring 2001.  Figure 1 shows a tentative detection of this feature in 
March 2000 and clear detections in January and December 2001 and April 2002.  
We did not detect the feature at a 3$\sigma$ level of 30 mJy (channel 
spacing: 1.11 km s$^{-1}$) in November 2002.  Velocities, determined by
Gaussian fit,
were 7565.0 $\pm$ 0.8 km s$^{-1}$ in March 2000, 7565.6 $\pm$ 0.5 km s$^{-1}$ 
in January 2001, 7568.6 $\pm$ 0.7 km s$^{-1}$ in December 2001, and (two lines) 
7567.6 $\pm$ 0.1 km s$^{-1}$ and 7612.1 $\pm$ 0.1 km s$^{-1}$ in April 2002.
The feature seen at 7567.6 km s$^{-1}$ in April 2002 drifted by 1.4 $\pm$ 0.3 
km s$^{-1}$ yr$^{-1}$ between March 2000 and April 2002.  In Figure 2 we 
reproduce the sensitive GBT spectrum from April 2002 showing a 1600 km s$^{-1}$
velocity range.

\subsection{TXS 2226-184}

TXS 2226-184 has the largest apparent isotropic luminosity of the known water 
masers, $\sim$ 6000 $\sollum$ \citep{koekemoer95}.  While the profile 
resembles that of NGC 1052 in breadth, the maser in TXS 2226-184
has been much less variable during the 7 years it has been monitored.
The emission spans $\sim$ 400 km s$^{-1}$.
The excellent spectra from February 1998 and April 2002 show that the maser 
profile has complex structure including a bifurcated peak and a tail of 
emission extending to the blue.  
These features may exist but be unseen in the less sensitive spectra.

The host galaxy, originally thought to be an elliptical, has recently been
observed with HST \citep{falcke00b}.  These authors argue that the
presence of an apparent dust lane and the shape of the isophotes are
indicative of a spiral host.

\subsection{IC 1481}

Maser emission near 6236 km s$^{-1}$ has been detected in IC 1481 
since the discovery of the maser in April 1995.  A second maser spike near 6251 
km s$^{-1}$ appeared in the July 1996 spectrum and has been detected 
through April 2002.  The velocities of these two maser peaks show
no discernible drift, with accelerations measured at 0.17 $\pm$ 0.12
km s$^{-1}$ yr$^{-1}$ (6236) and --0.08 $\pm$ 0.18 km s$^{-1}$ yr$^{-1}$ (6251).
These lines are redshifted from the systemic velocity, 6122 km s$^{-1}$. 
A flare appeared around 6310 km s$^{-1}$ on the 3 November 1997
spectrum, and was not seen on 12 December 1997 or afterwards.
\citet{sand99} observed this galaxy with the VLA
in order to measure the relative positions of the maser and continuum
emission.  They find a tentative offset of the maser from the continuum
of 14 $\pm$ 4 mas, a projected distance of 5 pc.  This extent is
significantly greater than that measured for other nuclear maser sources.
If we were to assume that the masers come from the edge of a disk which has
an orbital velocity of 120 km s$^{-1}$ (representative of the displacement
of the maser lines 
from the systemic velocity) then we can infer a black hole mass of 
1.7 $\times$ 10$^7 \solmass$.  We caution that this measurement is extremely
speculative, but it does demonstrate a powerful technique for measuring
black hole masses as more sensitive observations become available.

\section{Masers in Inclined Galaxies}

\citet{braatz97} show that water masers tend to be detected in those
Seyfert 2 or LINER galaxies which are highly inclined, and the 
reclassification of TXS 2226-184 as an
inclined disk galaxy by \citet{falcke00b} strengthens that trend.  
One reason for this relation might be that the orientation of the
nuclear disk is tied to that of the galactic disk so that maser detections,
which favor edge-on nuclear geometries, would also favor highly inclined 
galactic disks.  There is weak evidence that such a coupling may exist in 
late-type spirals, but not early-type \citep{wilson94}.  Another reason that
highly inclined galaxies appear to be favored might be that 
an appreciable fraction of the maser amplification arises from non-nuclear gas 
in the galactic disk.  This would explain not only the tendency to find
masers in highly inclined galaxies, but also the tendency we have found in this
paper for maser lines to remain at a constant velocity.
The acceleration of gas towards the galactic center would be very small for
masers in the galactic disk, so no velocity drift would be detectable.

The most successful searches for H$_2$O masers have been those which target
known AGNs, but AGN catalogs are deficient in highly inclined galaxies.
A survey of nearby, bright galaxies indicates that a large fraction (43\%)
have nuclear emission lines characteristic of an AGN, although their power is
generally weaker than ``classical'' AGNs \citep{ho97}.  We might
expect that a similar fraction of bright, highly inclined galaxies would
have some level of nuclear activity, but that such activity has gone 
undetected due to obscuration.  A luminous AGN is not a requirement for strong
maser emission, as exemplified by the strong maser emission in the 
low-luminosity AGN of NGC 4258.  Hence it is reasonable to search for water 
vapor masers in nearby, highly inclined galaxies even
if they are not known to be AGNs.  We searched 58 nearby galaxies with 
galactic disk inclinations $>$ 80$^\circ$.  Table 2 shows the source list.
Observations were made in February 1998 with the Effelsberg 100 m, and in 
August 1998 with the Parkes 64 m.  No maser sources were detected.
These observations suggest that clouds in the large scale galactic disks
do not contribute to the gain of masers originating in the nuclear region
and/or these galaxies do not possess nuclear activity required to produce
the megamaser.  We note that precise Hubble classifications are uncertain for 
many of the sample because of the high inclinations, but more than half are 
late-type galaxies, which are less likely to have nuclear activity.

\section{Discussion}

The spectral profile of water masers in NGC 4258 and the velocity changes
of the individual narrow features exemplify gas in a thin,
edge-on disk: the systemic features drift through a window of velocities
about 100 km s$^{-1}$ wide while masers associated with the projected 
extremities of the disk remain stationary.  NGC 2639 is the only source
revealed by this study to partly duplicate this pattern, and only then
in its systemic features; there is not yet a detection of high velocity lines
from this galaxy.  Indeed, H$_2$O maser lines in other galaxies more 
commonly have constant velocities,
regardless of whether they are near the systemic velocity of the 
galaxy or separated from it (by up to 120 km s$^{-1}$ in these cases).  The 
current emergence of much more sensitive instruments will be used to test
this trend further.  Individual systemic features in NGC 1386, for example, 
can be distinguished now with the GBT.

That we do not detect an acceleration
in most of the sources indicates that either the gravitational force on the 
maser clouds is weaker than that in NGC 4258, or that we are preferentially 
detecting masers whose gravitational acceleration vector is close to the 
plane of the sky, as is the case for the high velocity masers in NGC 4258.
If the masers amplify nuclear continuum emission, which originates nearer the
nucleus than the masers themselves, then the maser gas necessarily must
have a gravitational acceleration directed away from the observer,
regardless of the detailed geometry of the masing gas.  For the resulting
acceleration to go undetected by our observations, 
the gas producing systemic maser features would
have to be several times more distant from the central mass than in NGC 4258
(0.14 pc), assuming a similar black hole mass (3.9 $\times 10^7 \solmass$, 
\citet{herrnstein99}).
If, instead, the masers are self-amplified
structures analogous to the high velocity features in
NGC 4258, then the relatively small separations between maser velocities and 
systemic velocities (40 -- 120 km s$^{-1}$) indicate more slowly rotating 
structures than the 1000 km s$^{-1}$ rotating disk in NGC 4258.  This might
then suggest that there has been a strong selection effect in past surveys used
to find new masers, because they were limited in the bandwidth of the spectra.  
Typically 40 -- 64 MHz bands have been used, corresponding to a velocity
range of 540 -- 863 km s$^{-1}$.  We would only detect high velocity features
in those sources with
larger, slowly rotating nuclear disks.  The absence of systemic features 
in this case might be caused by the seed radio continuum source being very 
weak, absent, or offset from the plane of the maser disk.

\acknowledgements
We thank the telescope operators and staff at Effelsberg and Green Bank and 
the staff at Parkes for expert assistance throughout this project.  We thank 
Lincoln Greenhill and Jim Moran for stimulating discussions, and Ron Maddalena
for his talents with the GBT.
We have made extensive use of the NASA/IPAC Extragalactic
Database (NED) which is operated by the Jet Propulsion Laboratory,
California Institute of Technology, under contract with the National
Aeronautics and Space Administration.  This research was supported in part by
the National Science Foundation through grant AST 95-27289 to the
University of Maryland.

\clearpage

\begin{deluxetable}{lr}
\tablecaption{Dates of Observations}
\tablewidth{0pt}
\tablehead{\colhead{Telescope} & \colhead{Date}}

\startdata
Effelsberg & 15 -- 18 Oct 1993 \\ 
           & 31 Jan -- 1 Feb 1994 \\ 
           & 21 Mar -- 2 Apr 1994 \\ 
           &  4 -- 29 Apr 1995 \\ 
           & 25 -- 26 Jun 1995 \\ 
           & 15 -- 17 Sep 1995 \\ 
           & 13 -- 15 Jan 1996 \\ 
           & 22 -- 24 Dec 1996 \\ 
           &  4 --  9 Apr 1997 \\ 
           &  3 --  5 Nov 1997 \\ 
           &       12 Dec 1997 \\ 
           &  2 --  4 Jan 1998 \\ 
           & 30 Jan -- 11 Feb 1998 \\ 
	   & 27 -- 28 Oct 1999 \\ 
	   & 18 -- 19 Mar 2000 \\ 
	   & 16 -- 19 Jun 2000 \\ 
	   & 25 -- 29 Jan 2001 \\ 
	   &       30 Dec 2001 \\ 
	   &       26 Nov 2002 \\ 
Green Bank &  4 --  6 Jul 1994 \\ 
140 ft     & 17 -- 21 Nov 1994 \\ 
           &  9 -- 12 Aug 1995 \\ 
           & 14 -- 16 Nov 1995 \\ 
           & 20 -- 22 Dec 1995 \\ 
           & 11 -- 12 Jan 1996 \\ 
           & 24 -- 26 Mar 1996 \\ 
           & 31 Mar -- 5 Apr 1997 \\ 
           & 31 May -- 4 Jun 1997 \\ 
           & 23 -- 29 Dec 1997 \\ 
           & 13 -- 22 Feb 1999 \\ 
Parkes     & 7 -- 14 Jun 1994  \\ 
           & 21 -- 25 Oct 1995 \\ 
           & 25 -- 28 Jul 1996 \\ 
           & 13 -- 18 Aug 1997 \\ 
           & 28 Aug -- 1 Sep 1998 \\
GBT        & 25 -- 27 Apr 2002 \enddata
\end{deluxetable}

\clearpage

\begin{deluxetable}{lrrrrc}
\tablecaption{Galaxies Searched for H$_2$O Maser Emission}
\tablehead{\colhead{Source} & \colhead{$\alpha$$_{1950}$} &
\colhead{$\delta$$_{1950}$} & \colhead{V$_{center}$} &
\colhead{$\sigma$} & \colhead{Channel Spacing} \\
\colhead{} & \colhead{} & \colhead{} & \colhead{(km s$^{-1}$)} &
\colhead{(mJy)} & \colhead{(km s$^{-1}$)}
}

\startdata
 NGC 55       &  00 12 38.0 &  -39 29 53 &  -150 & 109 &  0.42 \\ 
 UGC 1281     &  01 46 38.9 &   32 20 33 &   163 &  35 &  0.66 \\ 
 NGC 779      &  01 57 11.9 &  -06 12 19 &  1100 & 130 &  0.42 \\ 
 UGC 2082     &  02 33 22.7 &   25 12 27 &   710 &  28 &  0.66 \\ 
 ESO 154-G023 &  02 55 22.1 &  -54 46 17 &   300 & 142 &  0.42 \\ 
 NGC 1332     &  03 24 03.8 &  -21 30 30 &  1300 & 128 &  0.42 \\ 
 NGC 1406     &  03 37 22.0 &  -31 29 00 &   700 & 127 &  0.42 \\ 
 NGC 1448     &  03 42 52.7 &  -44 48 05 &  1000 & 146 &  0.42 \\ 
 NGC 1515     &  04 02 50.0 &  -54 14 18 &   800 & 120 &  0.42 \\ 
 NGC 1596     &  04 26 32.0 &  -55 08 11 &  1200 &  77 &  0.42 \\ 
 NGC 1560     &  04 27 08.2 &   71 46 29 &   -28 &  23 &  0.66 \\ 
 NGC 2188     &  06 08 21.0 &  -34 05 41 &   750 & 131 &  0.42 \\ 
 NGC 2310     &  06 52 16.0 &  -40 47 53 &   850 & 129 &  0.42 \\ 
 NGC 2591     &  08 30 44.9 &   78 11 58 &  1331 &  30 &  0.66 \\ 
 NGC 2654     &  08 45 11.4 &   60 24 21 &  1382 &  30 &  0.66 \\ 
 NGC 2820     &  09 17 43.7 &   64 28 16 &  1576 &  29 &  0.66 \\ 
 NGC 3003     &  09 45 38.5 &   33 39 20 &  1480 &  31 &  0.66 \\ 
 NGC 3432     &  10 49 42.9 &   36 53 09 &   611 &  19 &  1.32 \\ 
 NGC 3600     &  11 13 06.6 &   41 51 57 &   719 &  17 &  1.32 \\ 
 NGC 3666     &  11 21 49.7 &   11 37 03 &  1067 &  15 &  1.32 \\ 
 NGC 3669     &  11 22 36.9 &   57 50 51 &  1940 &  17 &  1.32 \\ 
 NGC 3717     &  11 29 03.0 &  -30 01 58 &  1400 &  95 &  0.42 \\ 
 UCG 6534     &  11 30 30.6 &   63 33 23 &  1273 &  16 &  1.32 \\ 
 NGC 3877     &  11 43 29.0 &   47 46 21 &   894 &  16 &  1.32 \\ 
 NGC 3955     &  11 51 25.0 &  -22 53 12 &  1200 & 119 &  0.42 \\ 
 NGC 3972     &  11 53 10.0 &   55 35 48 &   848 &  18 &  1.32 \\ 
 NGC 4010     &  11 56 03.0 &   47 32 21 &   905 &  16 &  1.32 \\ 
 NGC 4096     &  12 03 28.5 &   47 45 21 &   577 &  13 &  1.32 \\ 
 NGC 4129     &  12 06 19.3 &  -08 45 30 &   900 & 117 &  0.42 \\ 
 NGC 4144     &  12 07 27.5 &   46 44 09 &   267 &  20 &  1.32 \\ 
 NGC 4157     &  12 08 34.6 &   50 45 51 &   771 &  16 &  1.32 \\ 
 NGC 4173     &  12 09 50.2 &   29 28 57 &  1127 &  17 &  1.32 \\ 
 NGC 4183     &  12 10 46.8 &   43 58 33 &   934 &  24 &  1.32 \\ 
 NGC 4206     &  12 12 43.7 &   13 18 10 &   701 &  21 &  1.32 \\ 
 NGC 4244     &  12 15 00.0 &   38 05 10 &   247 &  20 &  1.32 \\ 
 NGC 4256     &  12 16 21.9 &   66 10 37 &  2531 &  19 &  1.32 \\ 
 NGC 4302     &  12 19 10.2 &   14 52 43 &  1118 &  20 &  1.32 \\ 
 NGC 4307     &  12 19 32.4 &   09 19 17 &  1305 &  24 &  1.32 \\ 
 NGC 4517     &  12 30 11.9 &   00 23 32 &  1131 &  24 &  1.32 \\ 
 NGC 4522     &  12 31 07.8 &   09 27 02 &  2331 &  22 &  1.32 \\ 
 NGC 4656     &  12 41 31.8 &   32 26 30 &   649 &  20 &  1.32 \\ 
 NGC 4845     &  12 55 28.1 &   01 50 42 &  1228 &  24 &  1.32 \\ 
 NGC 5023     &  13 09 58.1 &   44 18 15 &   400 &  27 &  1.32 \\ 
 NGC 5084     &  13 17 34.0 &  -21 33 56 &  1450 & 138 &  0.42 \\ 
 NGC 5290     &  13 43 11.6 &   41 57 45 &  2579 &  22 &  1.32 \\ 
 NGC 5301     &  13 44 21.4 &   46 21 25 &  1562 &  20 &  1.32 \\ 
 NGC 5529     &  14 13 27.7 &   36 27 27 &  2878 &  30 &  1.32 \\ 
 NGC 5775     &  14 51 26.8 &   03 44 51 &  1582 &  25 &  1.32 \\ 
 NGC 5894     &  15 10 32.7 &   59 59 38 &  2485 &  50 &  1.32 \\ 
 NGC 5984     &  15 40 33.4 &   14 23 25 &  1118 &  22 &  1.32 \\ 
 NGC 6368     &  17 24 51.2 &   11 35 01 &  2767 &  18 &  1.32 \\ 
 NGC 6810     &  19 39 21.0 &  -58 46 29 &  1730 & 170 &  0.42 \\ 
 NGC 6925     &  20 31 14.0 &  -32 09 12 &  2500 & 163 &  0.42 \\ 
 ESO 287-G013 &  21 19 56.0 &  -45 59 12 &  2715 & 121 &  0.42 \\ 
 NGC 7064     &  21 25 34.6 &  -52 59 11 &   800 & 132 &  0.42 \\ 
 NGC 7090     &  21 32 59.0 &  -54 46 52 &   857 & 133 &  0.42 \\ 
 NGC 7184     &  21 59 53.3 &  -21 03 17 &  2617 & 192 &  0.42 \\ 
 NGC 7361     &  22 39 31.0 &  -30 19 12 &  1245 & 185 &  0.42 \enddata
\tablecomments{The sample of highly inclined galaxies which was searched 
for H$_2$O emission.  No H$_2$O emission was detected.  The columns
show (1) galaxy name, (2) RA, (3) Declination, (4) Velocity (LSR) at the 
center of the
searched band (km s$^{-1}$), (5) 1-$\sigma$ noise value per channel (mJy), and 
(6) the channel spacing (km s$^{-1}$).  The galaxies listed with channel 
spacings of 0.66 or 1.32 km s$^{-1}$ were observed with the Effelsberg 100 m
with a 50 MHz bandwidth, and those listed with a 0.42 km s$^{-1}$ channel
spacing were observed with the Parkes 64 m with a 64 MHz bandwidth.}
\end{deluxetable}

\clearpage

\begin{figure}
\plotone{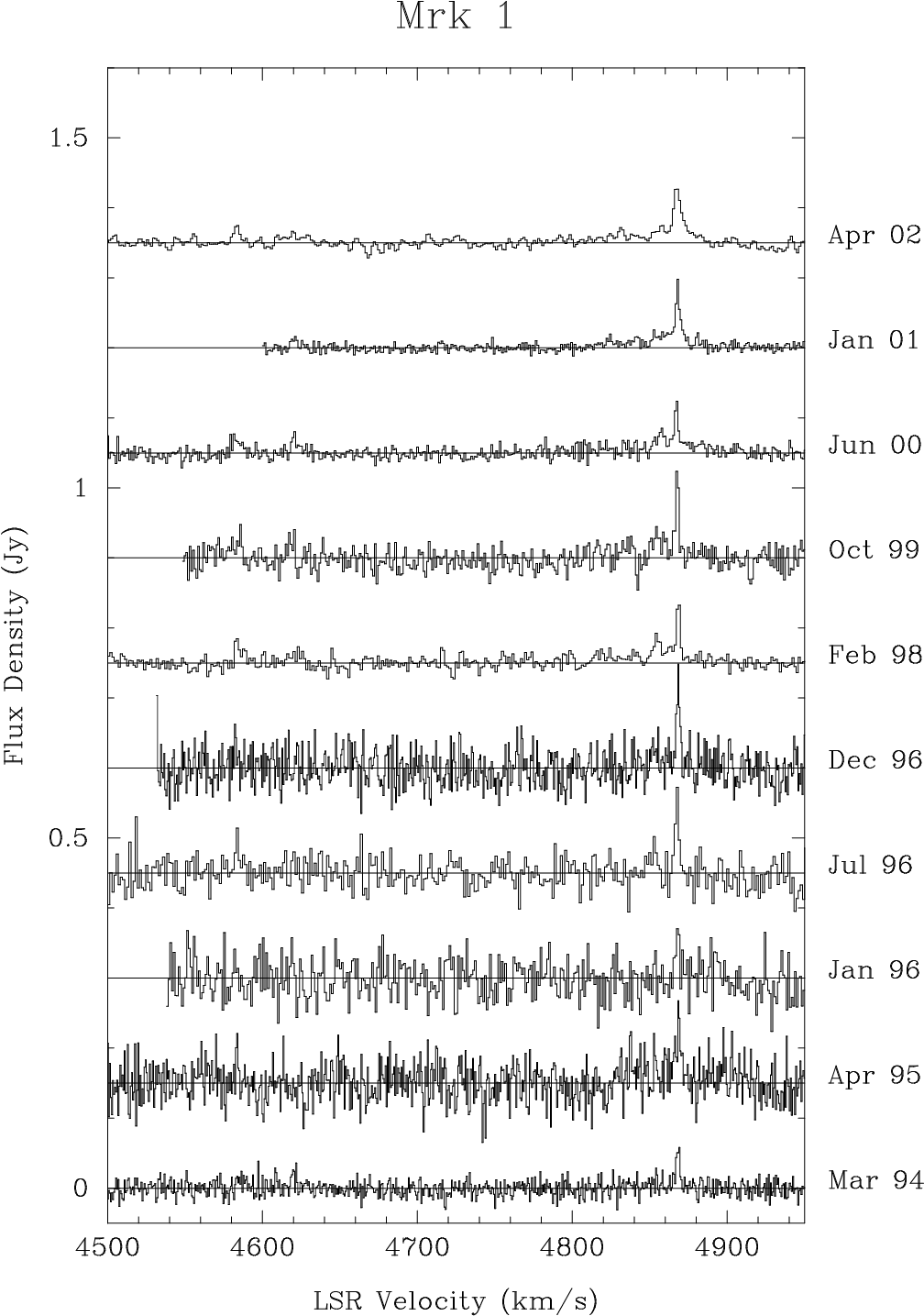}
\end{figure}
\clearpage
\begin{figure}
\plotone{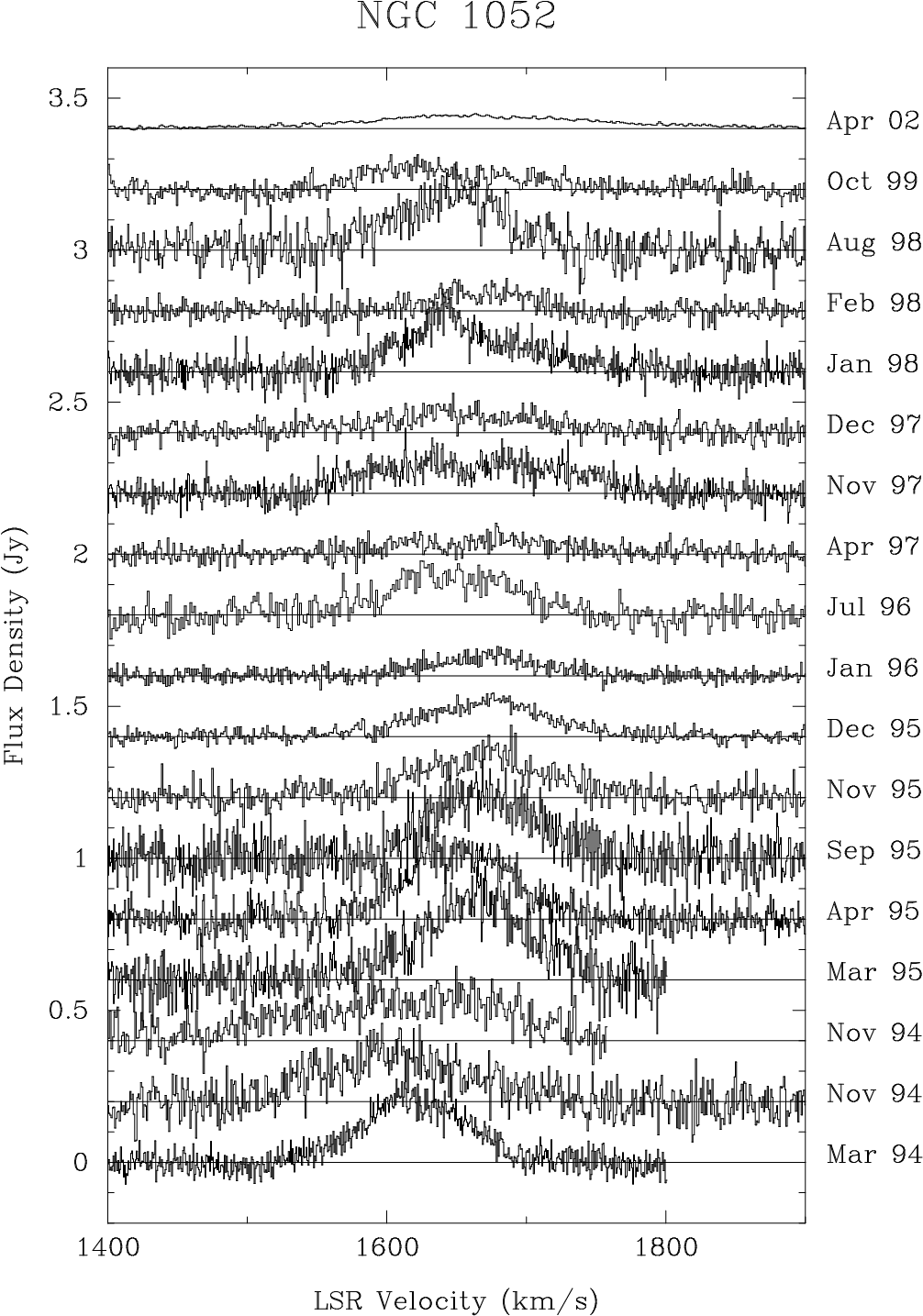}
\end{figure}
\clearpage
\begin{figure}
\plotone{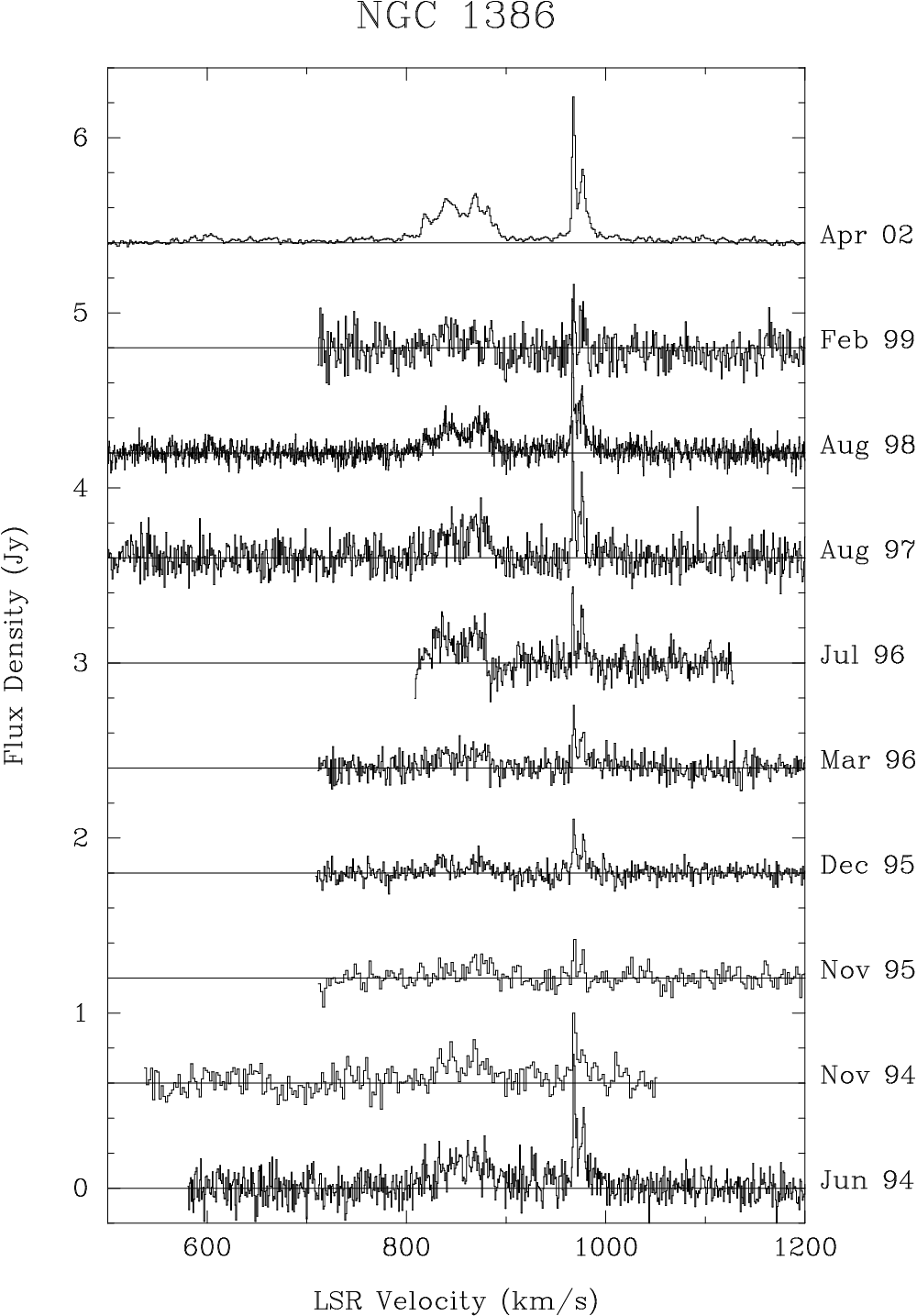}
\end{figure}
\clearpage
\begin{figure}
\plotone{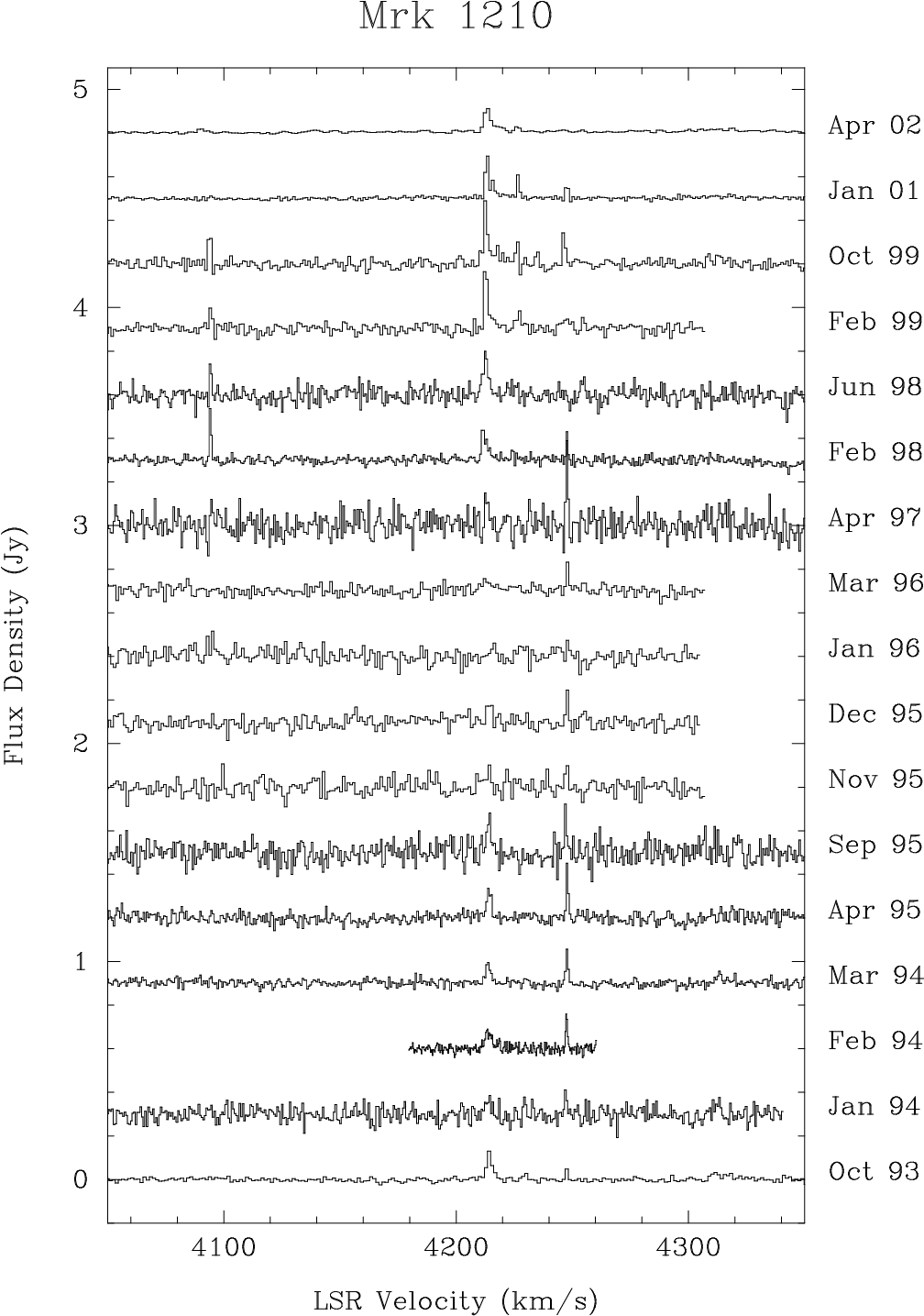}
\end{figure}
\clearpage
\begin{figure}
\plotone{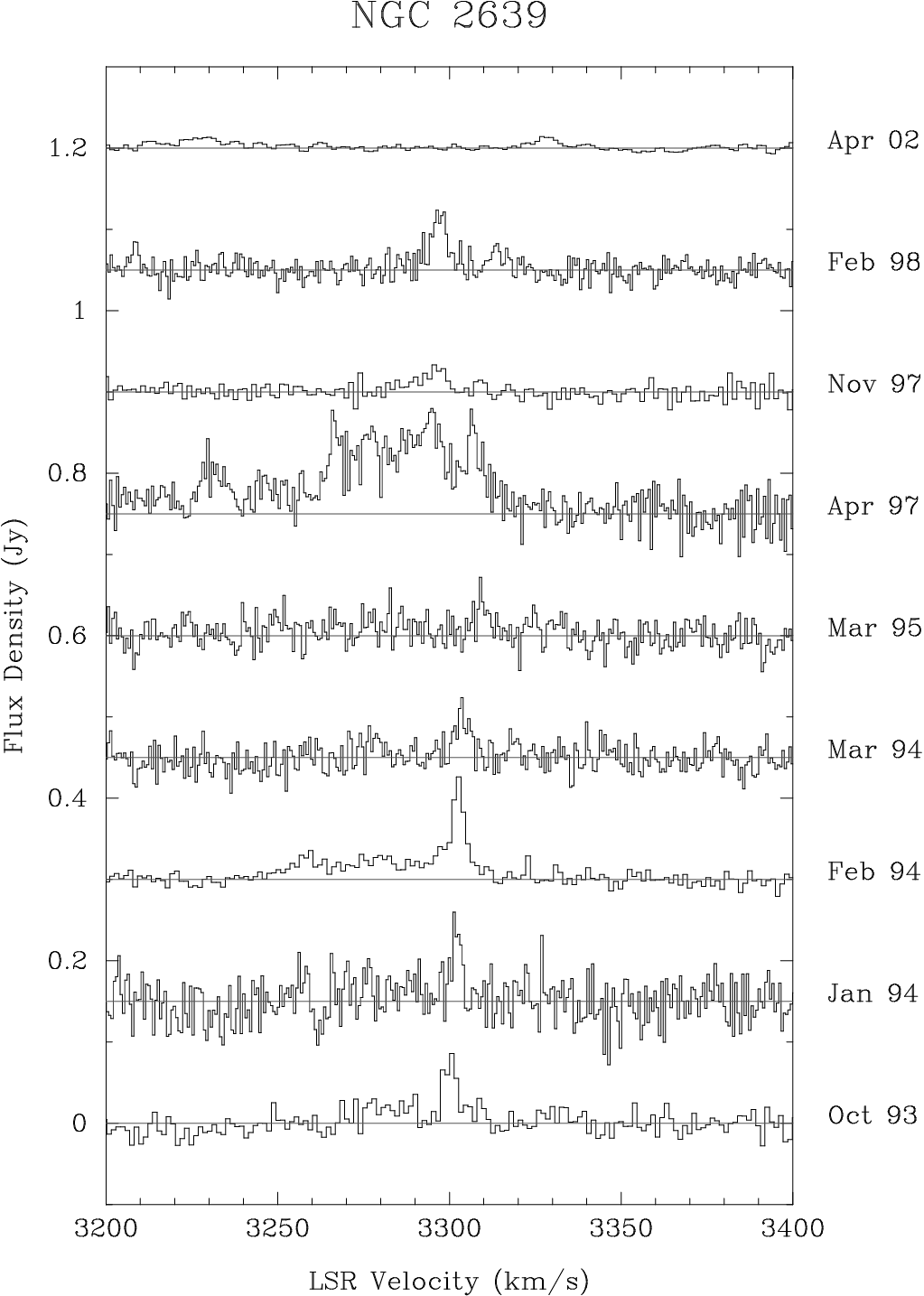}
\end{figure}
\clearpage
\begin{figure}
\plotone{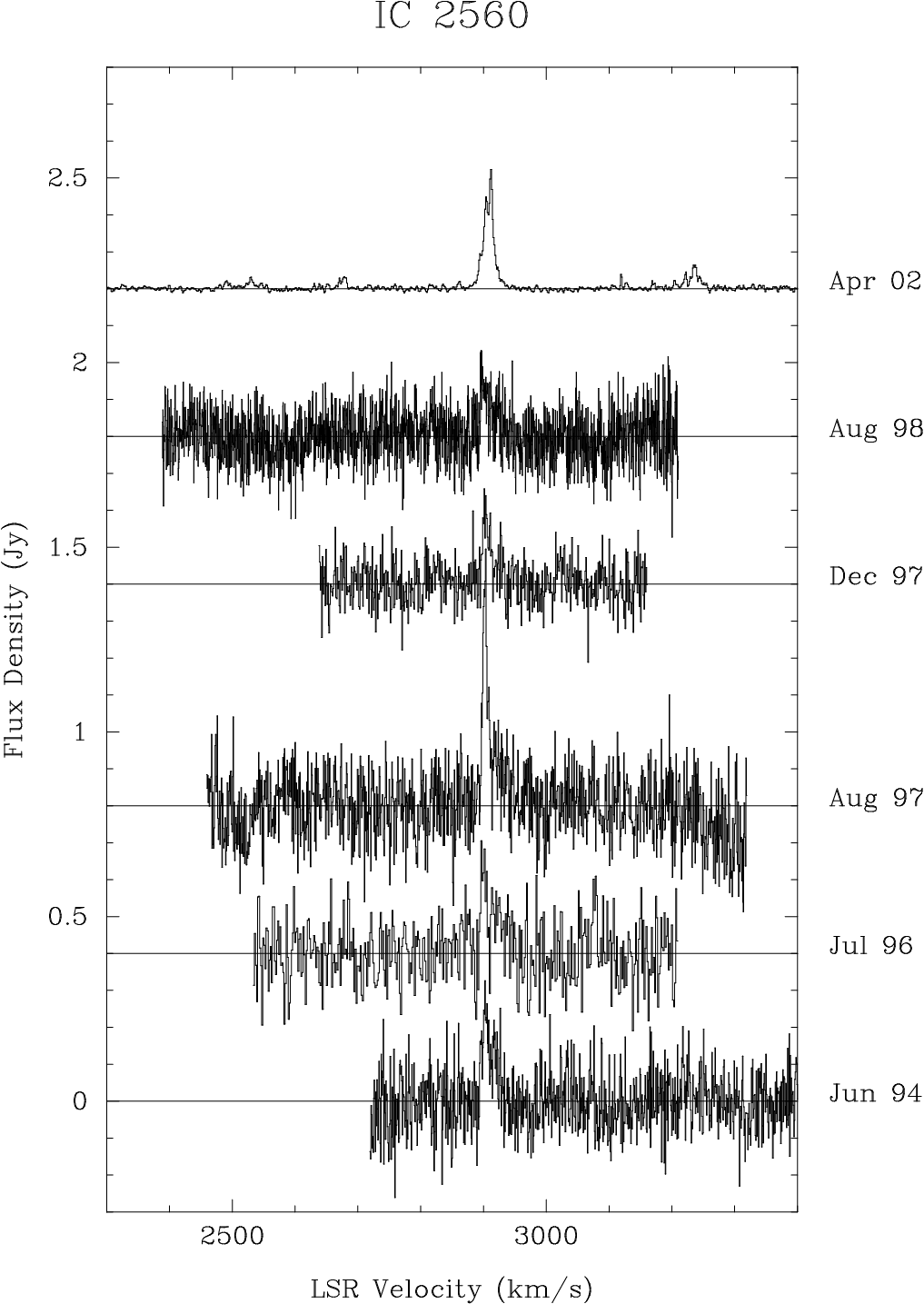}
\end{figure}
\clearpage
\begin{figure}
\plotone{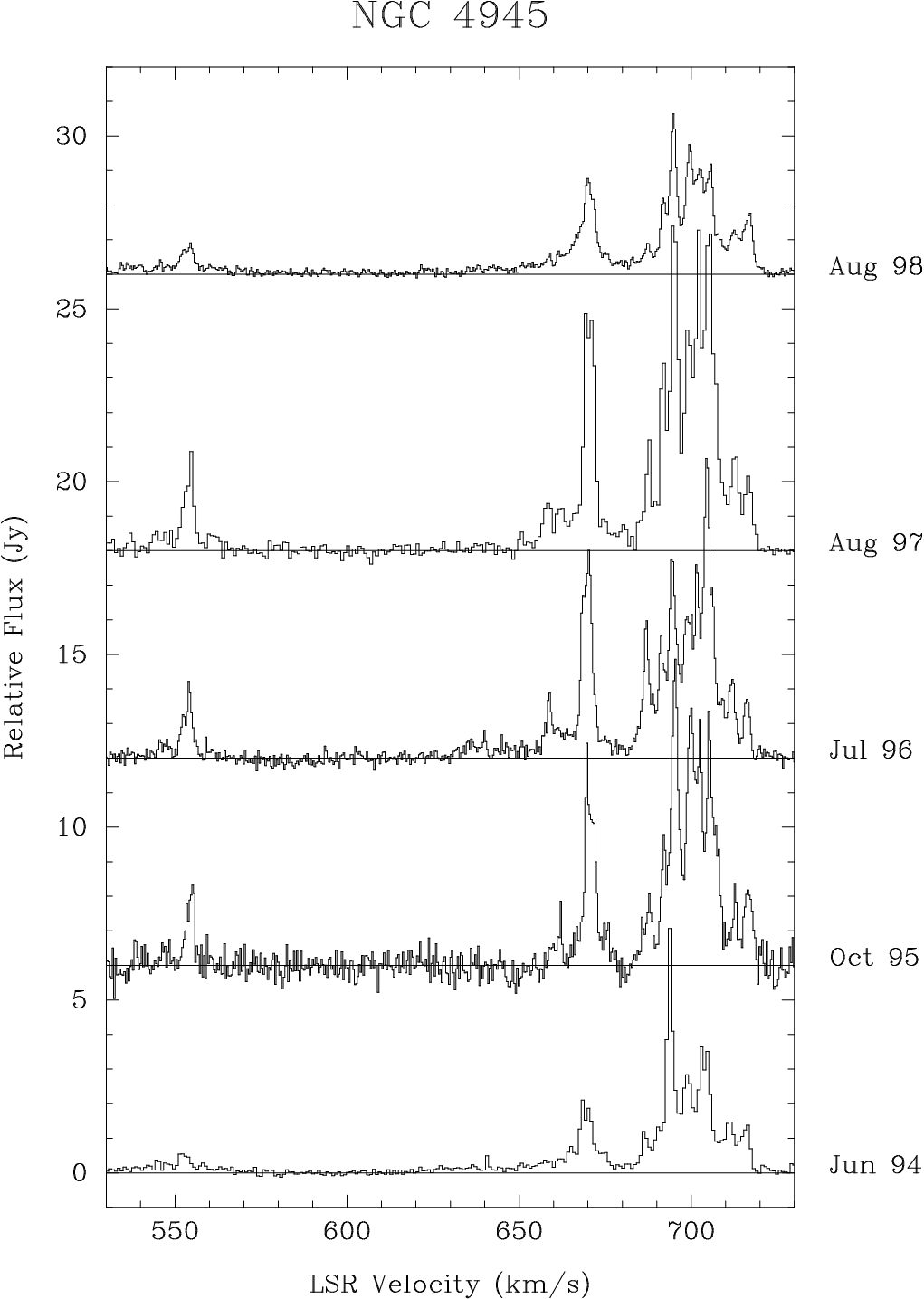}
\end{figure}
\clearpage
\begin{figure}
\plotone{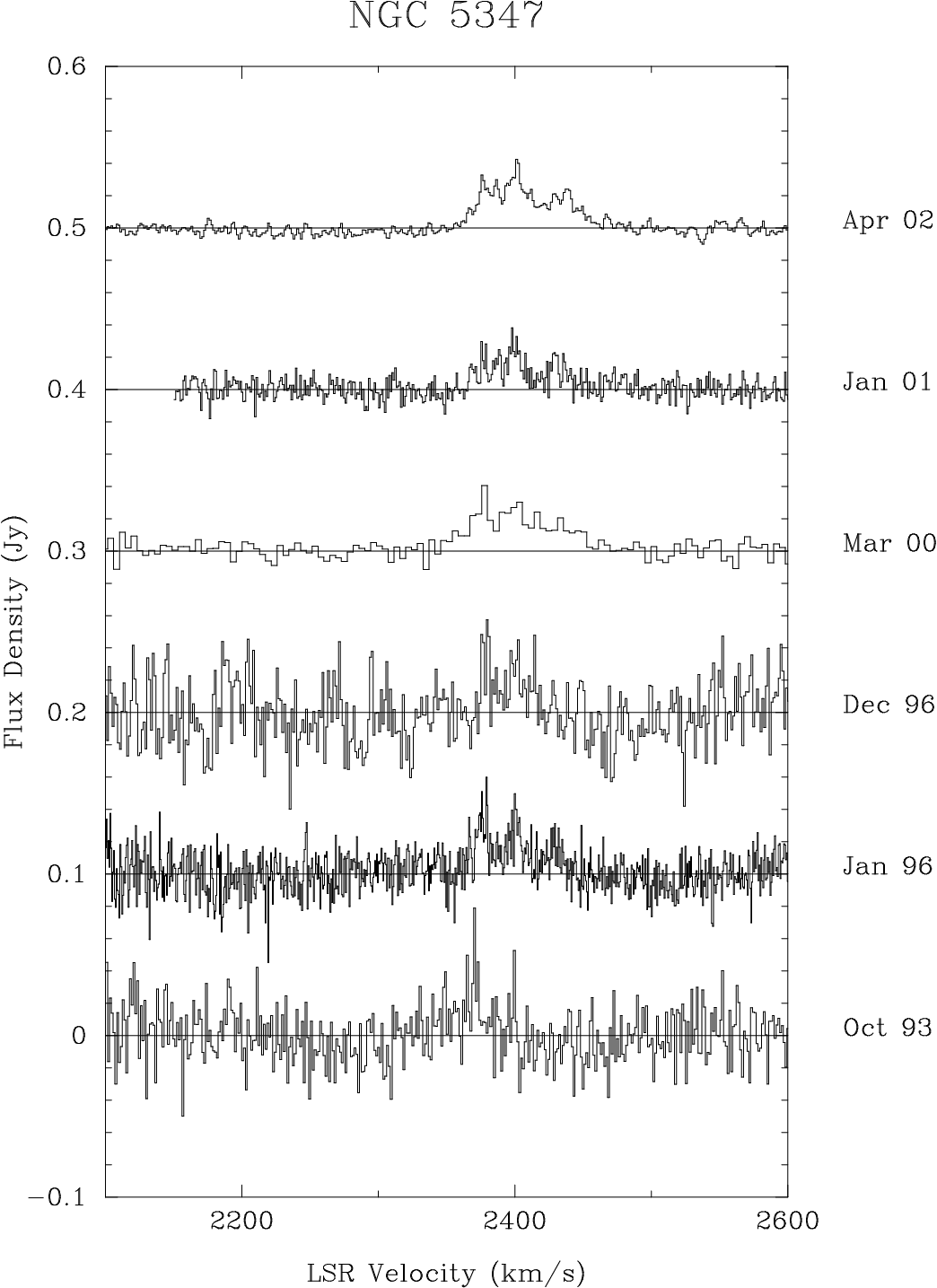}
\end{figure}
\clearpage
\begin{figure}
\plotone{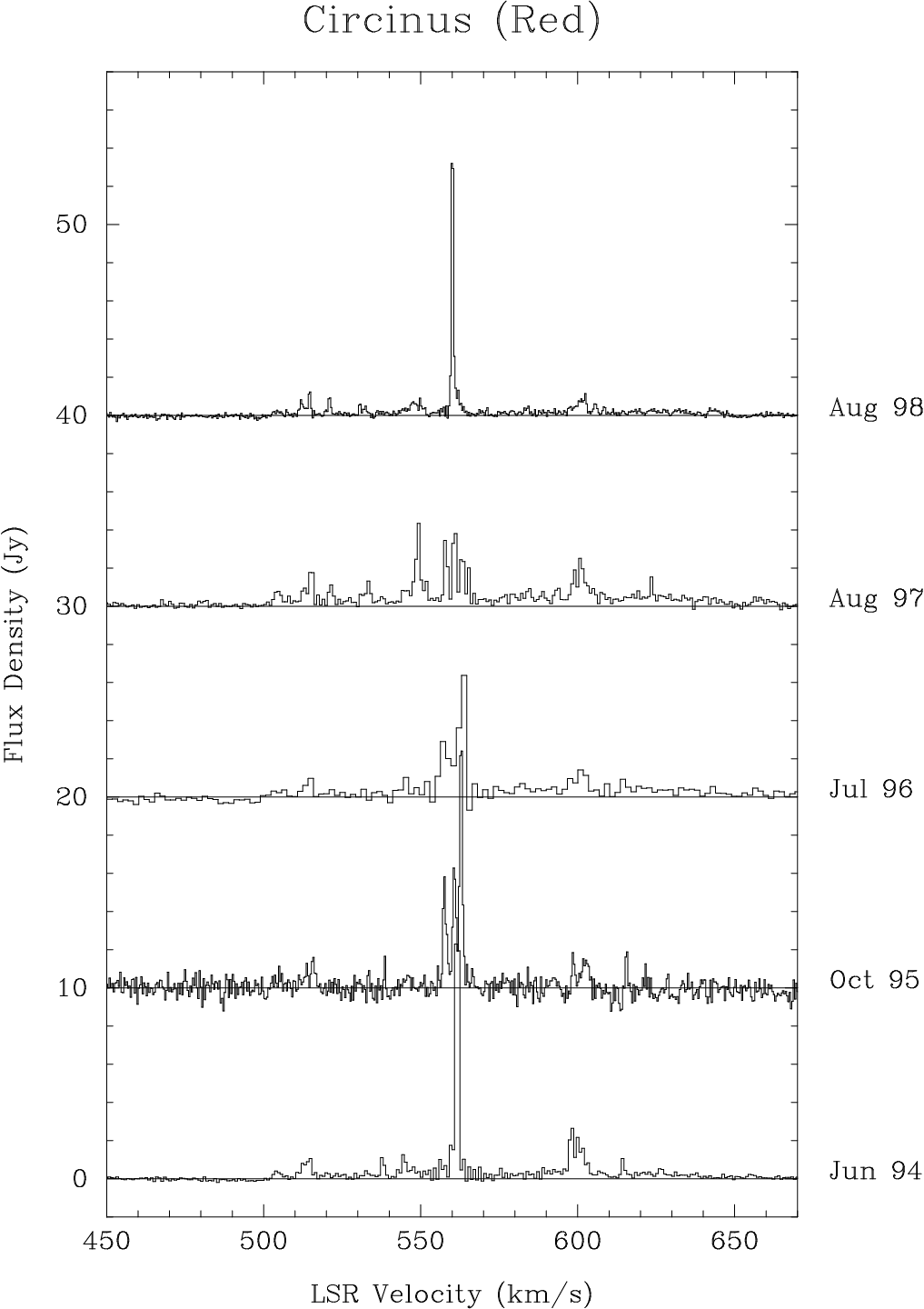}
\end{figure}
\clearpage
\begin{figure}
\plotone{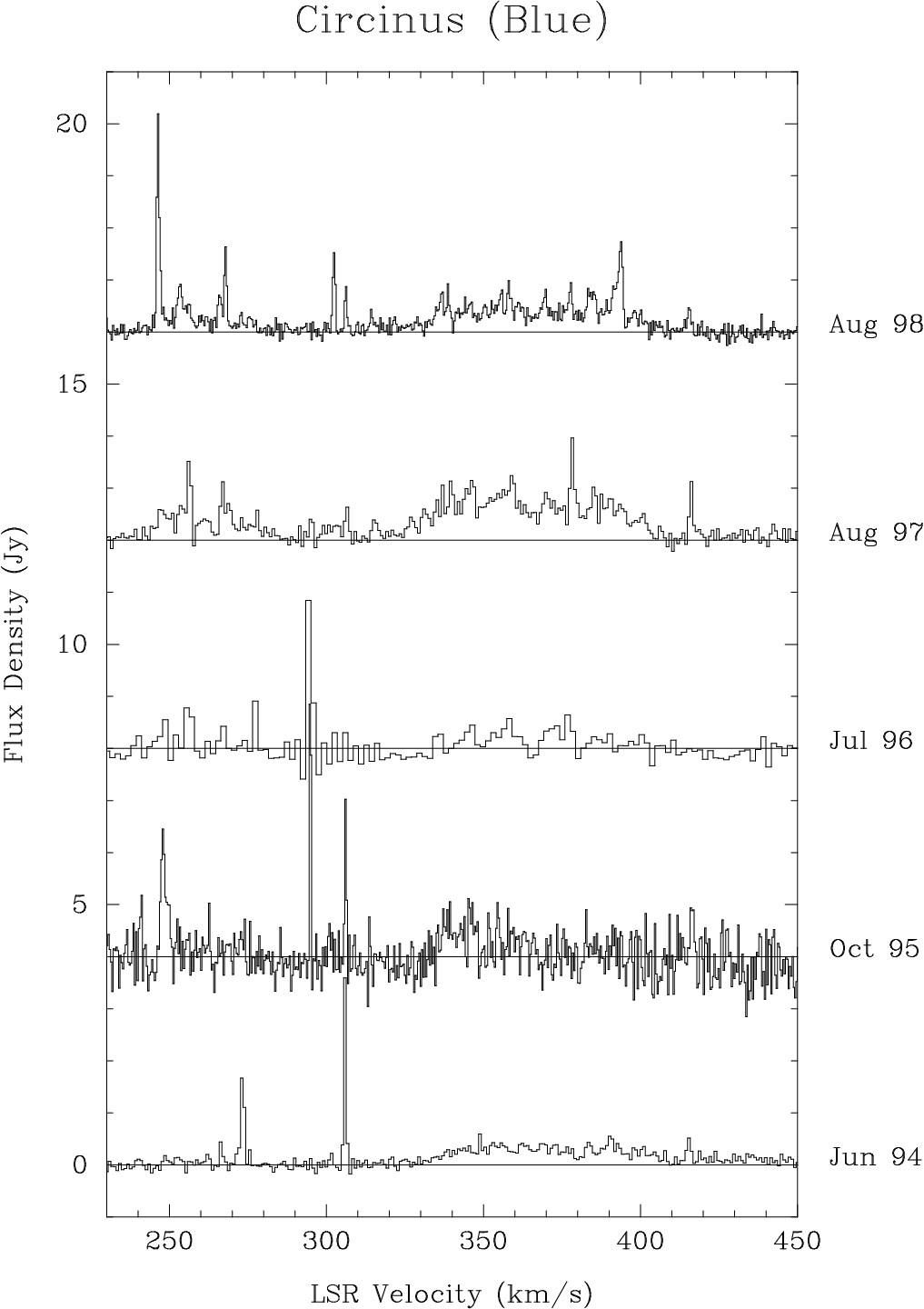}
\end{figure}
\clearpage
\begin{figure}
\plotone{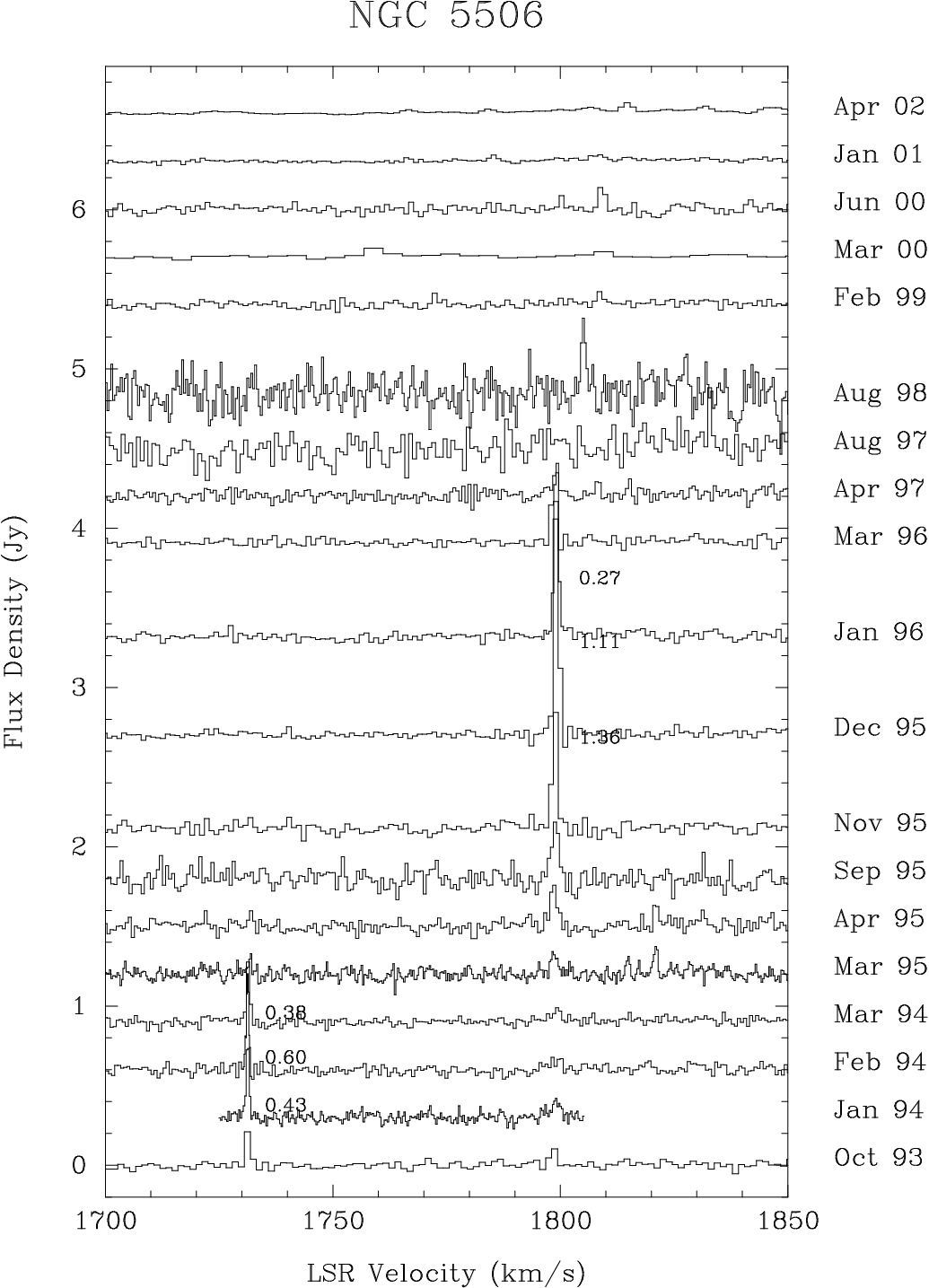}
\end{figure}
\clearpage
\begin{figure}
\plotone{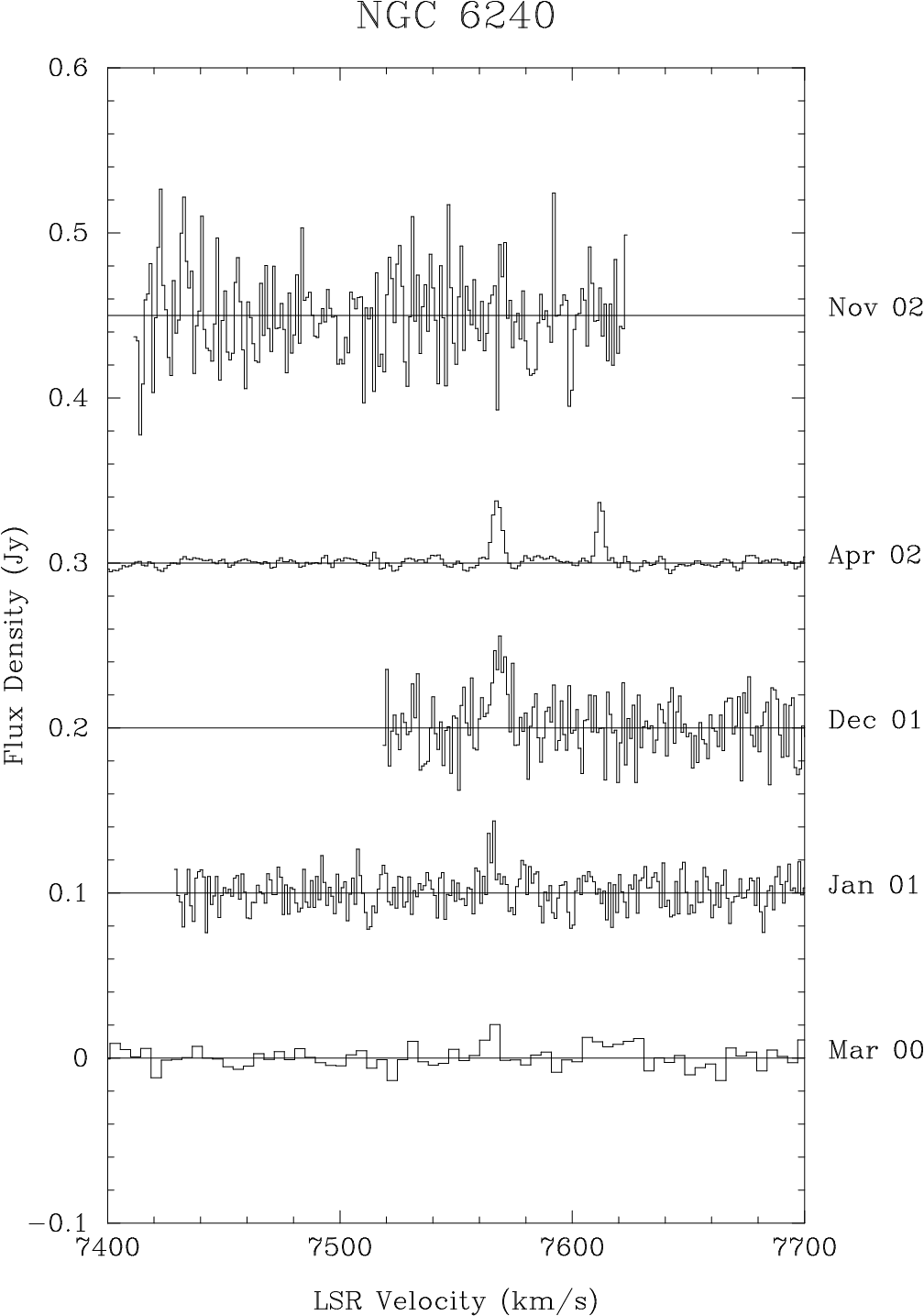}
\end{figure}
\clearpage
\begin{figure}
\plotone{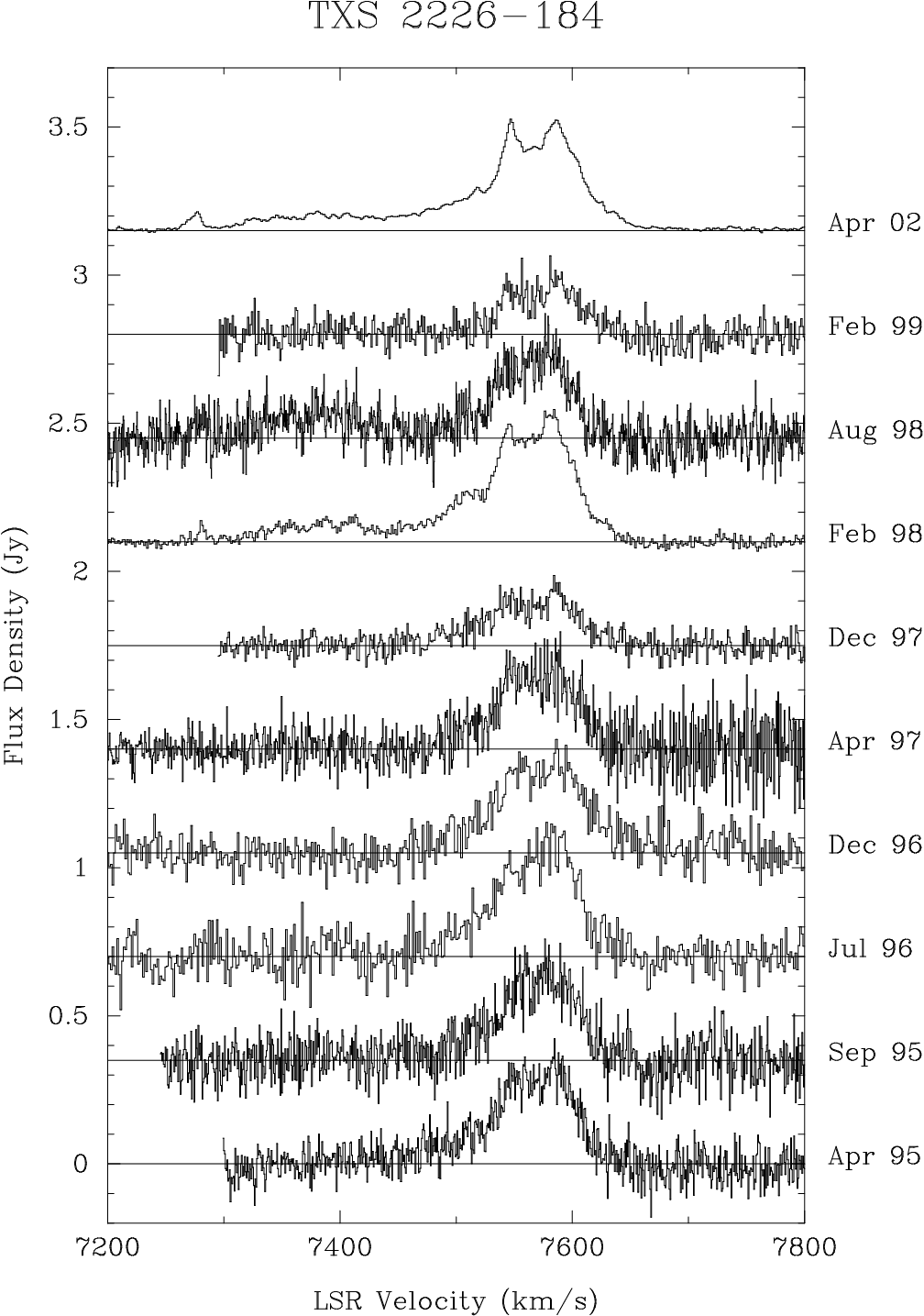}
\end{figure}
\clearpage
\begin{figure}
\plotone{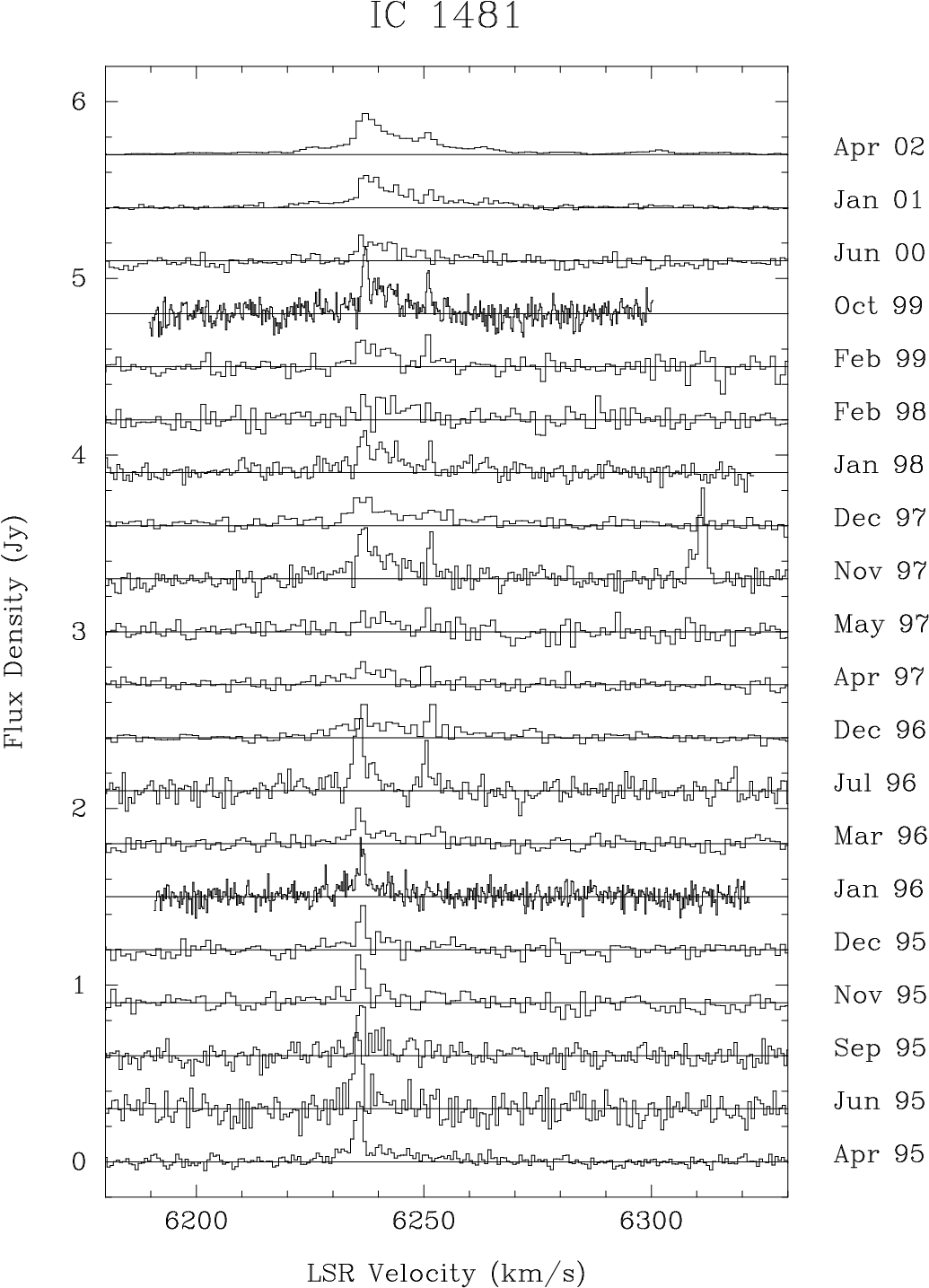}
\end{figure}
\clearpage
\begin{figure}
\figurenum{1}
\caption{Spectra of 1.3 cm water maser emission towards the nuclei
of 13 galaxies.  The velocity ranges shown for each source are chosen to 
facilitate comparison of the maser features, and so are not necessarily 
centered at the systemic velocity of the galaxy.  Velocities are
measured with respect to the LSR and use the optical convention.}
\end{figure}
\clearpage
\begin{figure}
\plotone{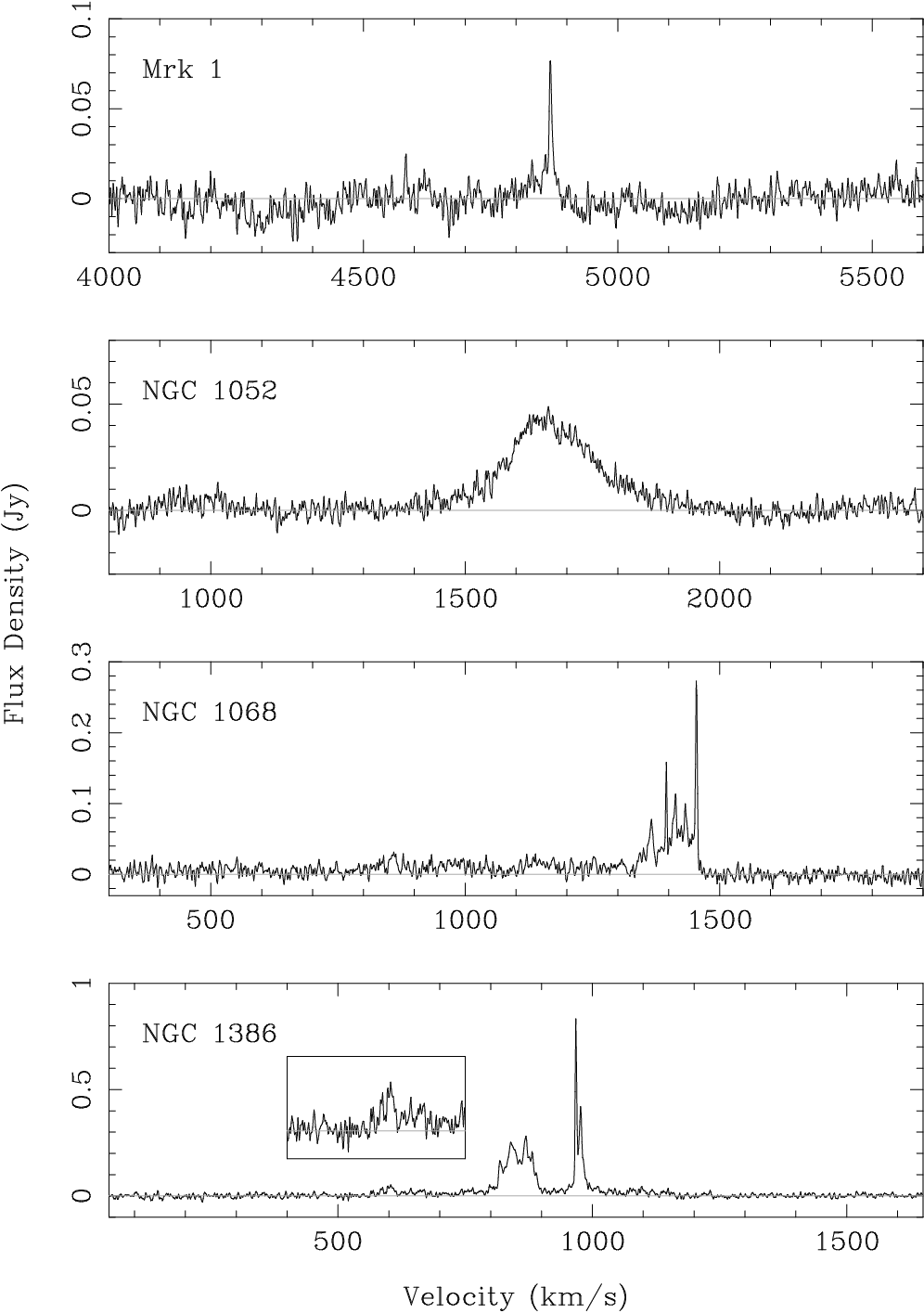}
\end{figure}
\clearpage
\begin{figure}
\plotone{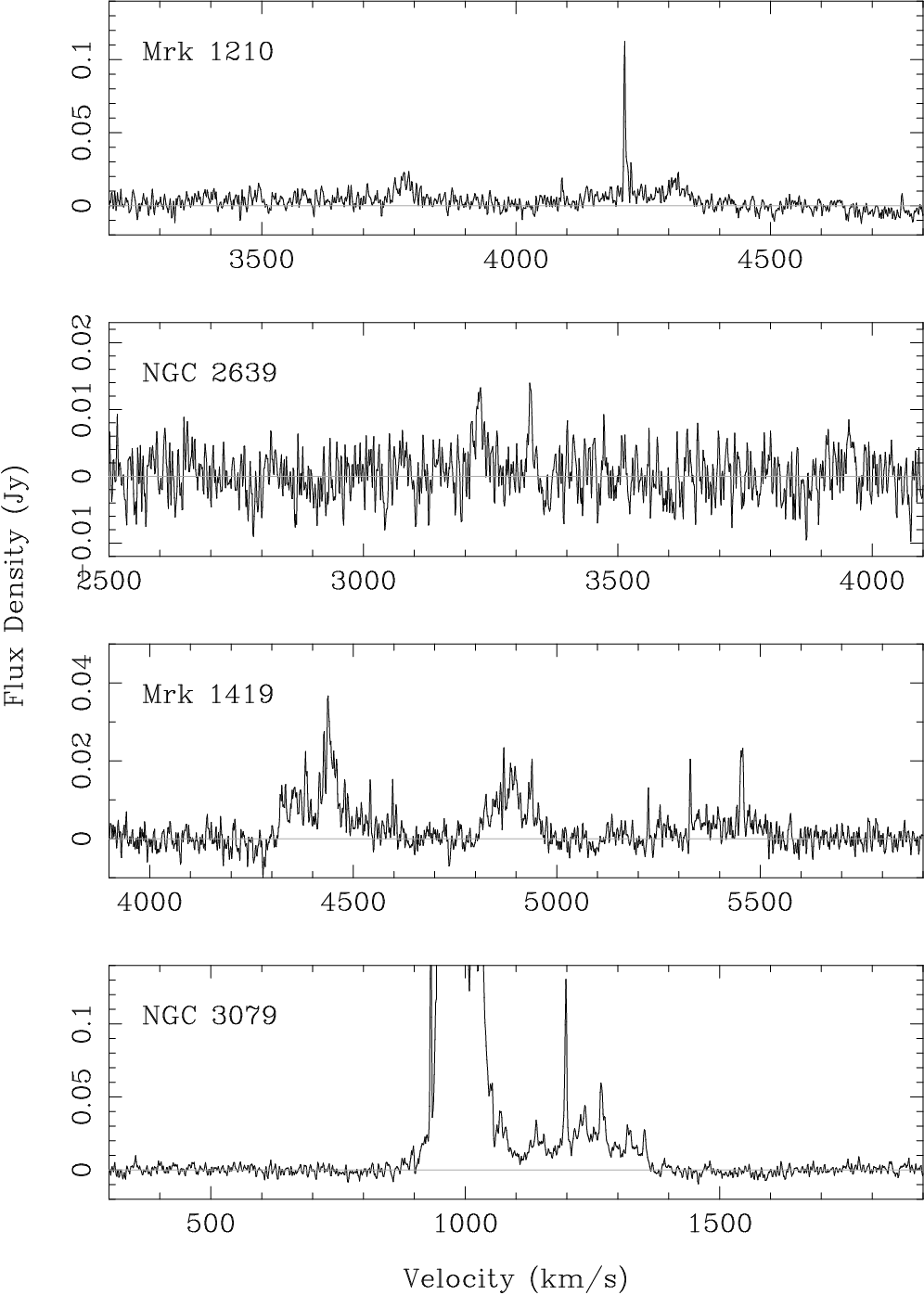}
\end{figure}
\clearpage
\begin{figure}
\plotone{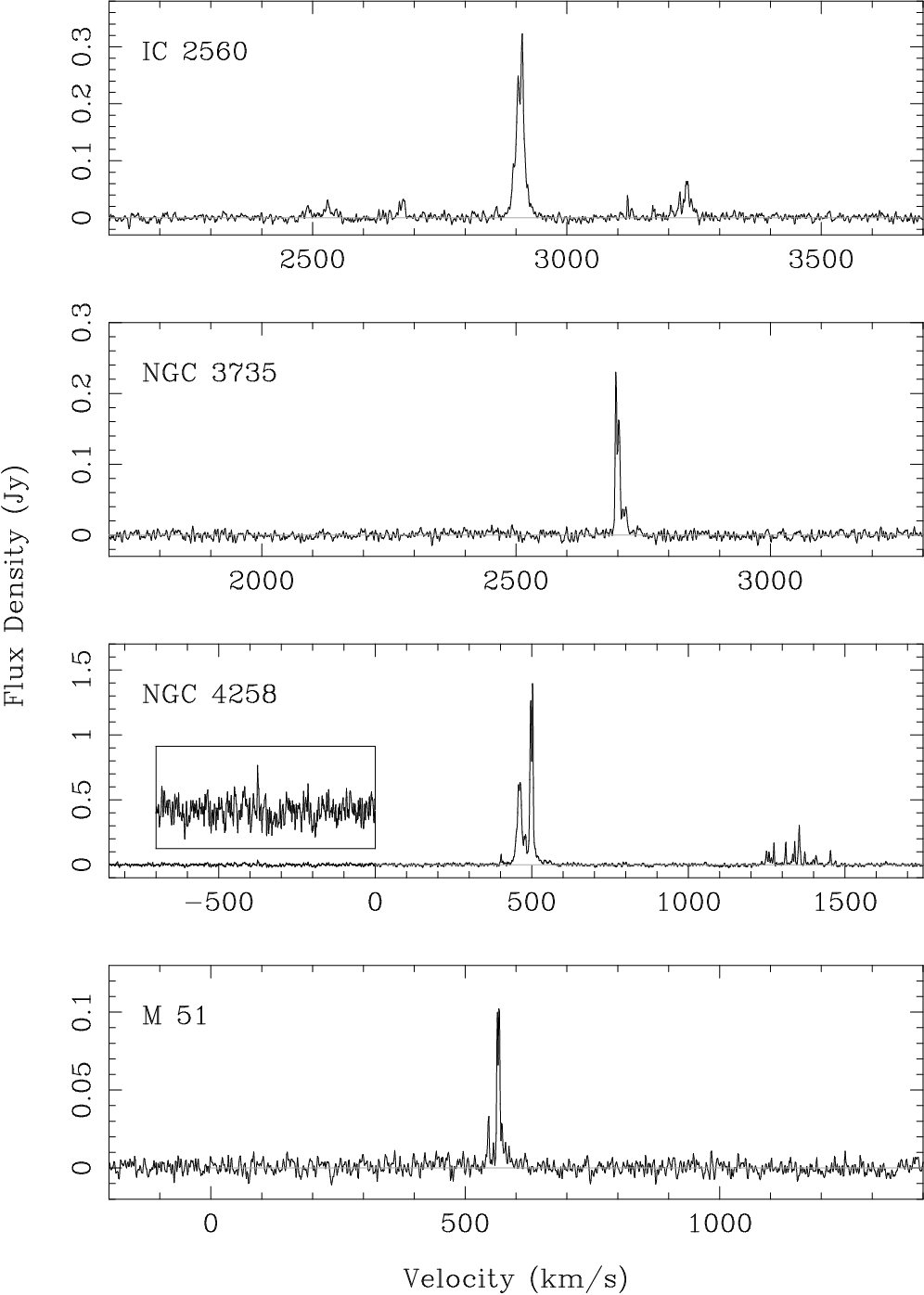}
\end{figure}
\clearpage
\begin{figure}
\plotone{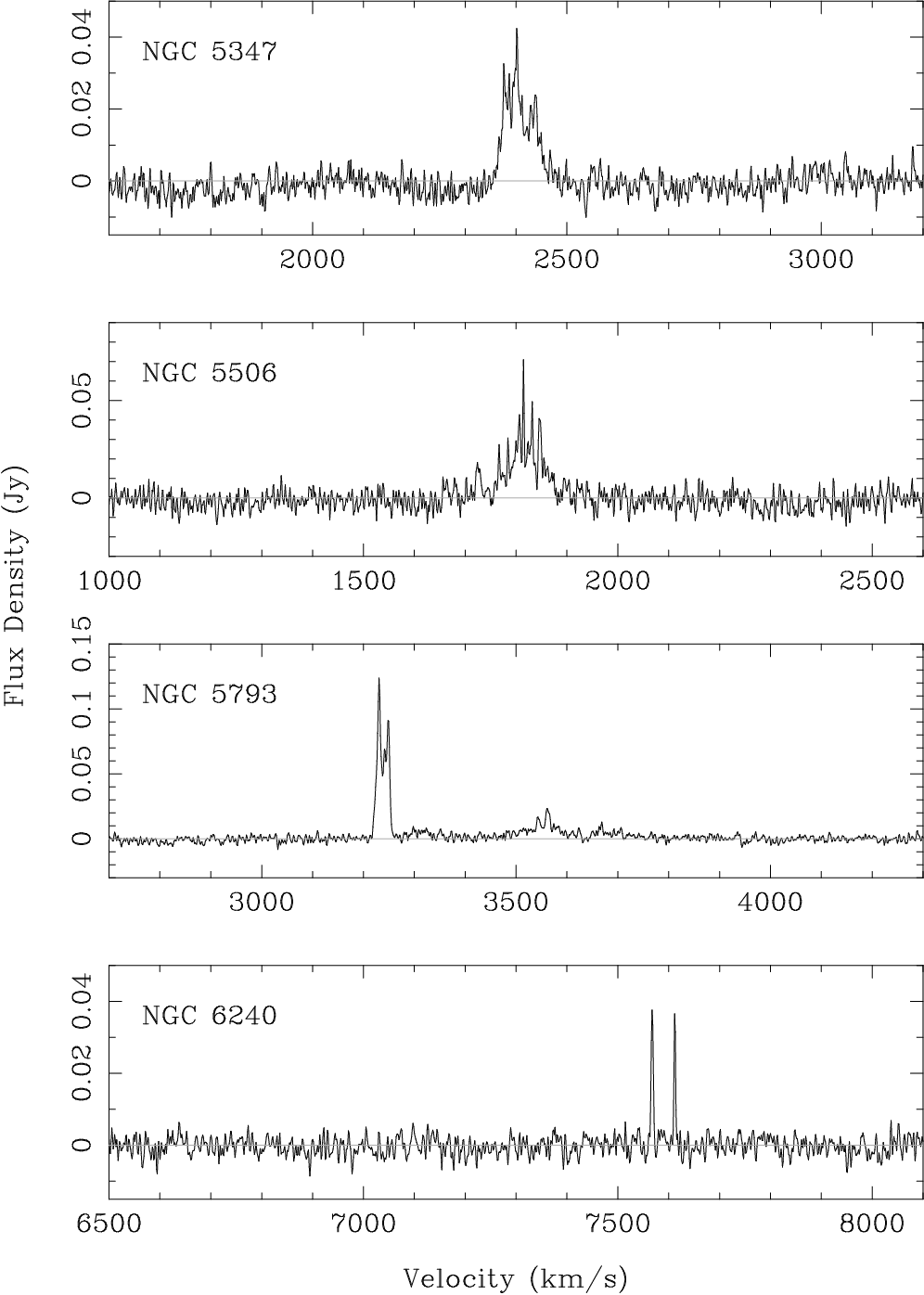}
\end{figure}
\begin{figure}
\figurenum{2}
\plotone{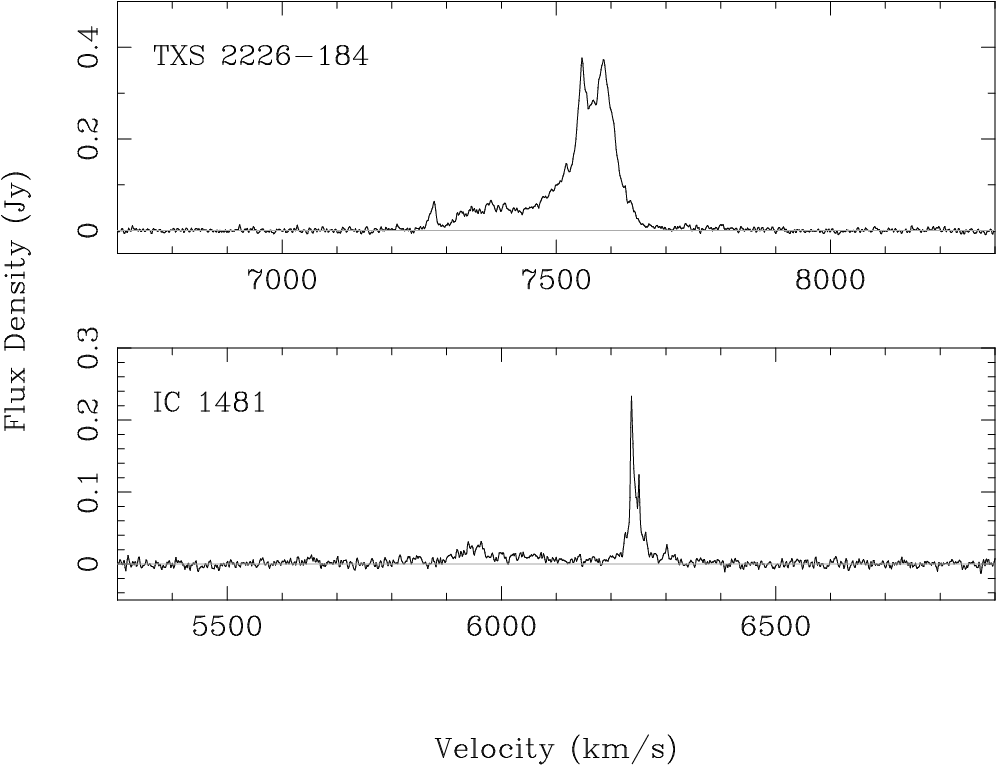}
\caption{Spectra of 18 maser sources observed with the GBT.  These represent, 
in most cases, the most sensitive spectra taken at a single epoch for each 
galaxy.  The spectra are centered approximately on the systemic recession 
velocity of the galaxy, and cover a velocity range of 1600 km s$^{-1}$ in each 
case except Mrk 1419 (which covers 2000 km s$^{-1}$) and NGC 4258 (which 
covers 2600 km s$^{-1}$). The flux scale for NGC 3079 is compressed to 
emphasize the faint features.  Velocities are measured with respect to the LSR 
and use the optical convention.}
\end{figure}
\end{document}